\documentclass[sigconf]{acmart}

\AtBeginDocument{%
  \providecommand\BibTeX{{%
    \normalfont B\kern-0.5em{\scshape i\kern-0.25em b}\kern-0.8em\TeX}}}

\usepackage{enumitem}
\usepackage{graphics}
\usepackage{xspace}
\usepackage{soul}
\usepackage[title]{appendix}
\usepackage{xcolor}
\usepackage{color, colortbl, soul}
\usepackage{subfigure}

\def\ie{\textit{i.e.,}\xspace}
\def\etal{\textit{et al.}\xspace}

\def\eg{\textit{e.g.,}\xspace}

\def\vs{\textit{vs.}\xspace}

\def\posthoc{\textit{post-hoc}\xspace}

\copyrightyear{2023} 
\acmYear{2023} 
\setcopyright{rightsretained} 
\acmConference[ASSETS '23]{The 25th International ACM SIGACCESS Conference on Computers and Accessibility}{October 22--25, 2023}{New York, NY, USA}
\acmBooktitle{The 25th International ACM SIGACCESS Conference on Computers and Accessibility (ASSETS '23), October 22--25, 2023, New York, NY, USA}
\acmDOI{10.1145/3597638.3608378}
\acmISBN{979-8-4007-0220-4/23/10}


\begin{teaserfigure}
    \centering
    \includegraphics[width=\textwidth]{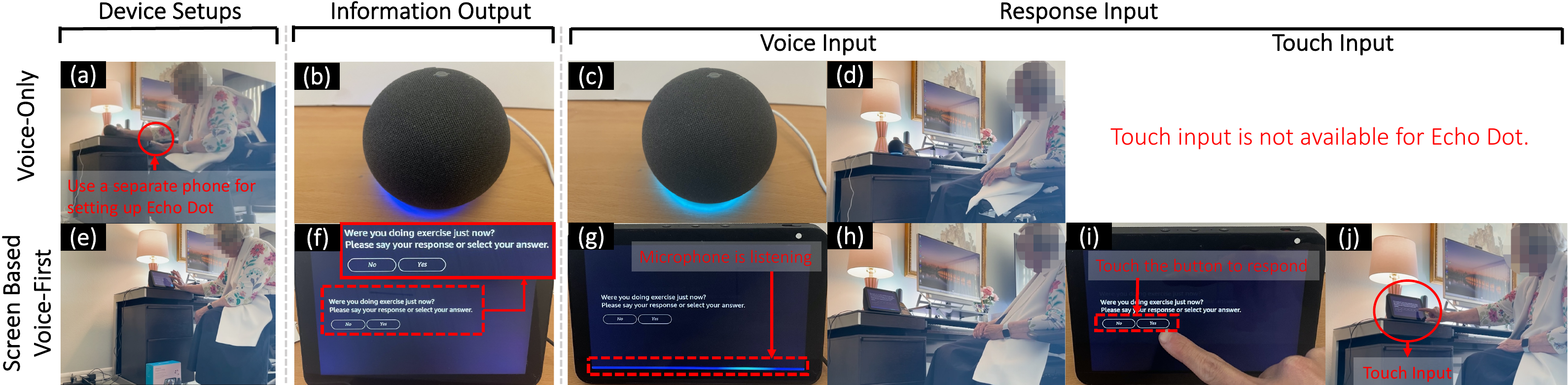}
    \caption{Our real-world deployment study aims to understand the affordances brought by the built-in touchscreen of Voice Assistants (VAs). We focus on device setup phases (a, e), as well as long-term uses for conducting diary survey and other miscellaneous general purposes. Notably, the touchscreen-based voice-first VAs allow older adults to \textit{see} prompts (f) and input responses by either \textit{voice} \mbox{(g - h)} or \textit{touch} (i - j), compared to the voice-only VAs (b - d).}
    \Description{Figure 1a shows that an older adult is using a separate mobile phone to setup the Echo Dot. Figure 1b shows an Echo Dot is announcing voice information. The bottom LED rings are shown as blue. Figure 1c shows an Echo Dot is listening to the user's speech. The bottom LED rings are shown as cyan. Figure 1d shows an older adult is speaking toward the Echo Dot. Figure 1e shows an older adult is setting up and configuring the Echo Show. Figure 1f demonstrates that the Echo Show is announcing the prompt of the designed daily diary. The question ``were you doing exercise just now? Please say your response or select your answer'' is shown on the screen. Two buttons — ``Yes'' and ``No'' are displayed where the user could input their responses by touching the display. Figure 1g shows the Echo Show, where the blue/cyan bar is displayed at the bottom of the screen. Figure 1h shows that an older adult is speaking toward the Echo Show. Figure 1i shows that an older adult is attempting to touch ``Yes'' of the question shown in Figure 1f. Figure 1j demonstrates that an older adult is attempting to input her response by touching the virtual button displayed on the Echo Show.}
    \label{fig::teaser}
\end{teaserfigure}


\begin{document}

\title[Screen or No Screen?]{Screen or No Screen? Lessons Learnt from a Real-World Deployment Study of Using Voice Assistants With and Without Touchscreen for Older Adults}

\author{Chen Chen}
\orcid{0000-0001-7179-0861}
\affiliation{%
  \department{Computer Science and Engineering}
  \institution{University of California San Diego}
  \city{La Jolla}
  \state{CA}
  \country{United States}
}
\email{chenchen@ucsd.edu}

\author{Ella T. Lifset}
\orcid{0000-0003-2640-3206}
\affiliation{%
  \department{Biological Sciences}
  \institution{University of California San Diego}
  \city{La Jolla}
  \state{CA}
  \country{United States}
}
\email{etlifset@ucsd.edu}

\author{Yichen Han}
\authornote{The author contributed to the project while at the University of California San Diego.}
\orcid{0000-0002-2997-2140}
\affiliation{%
  \department{Electrical and Computer Engineering}
  \institution{Carnegie Mellon University}
  \city{Pittsburgh}
  \state{PA}
  \country{United States}
}
\email{yichenha@andrew.cmu.edu}

\author{Arkajyoti Roy}
\orcid{0000-0003-0597-6582}
\affiliation{%
  \department{Department of Mathematics}
  \institution{University of California San Diego}
  \city{La Jolla}
  \state{CA}
  \country{United States}
}
\email{aroy@ucsd.edu}

\author{Michael Hogarth}
\orcid{0000-0002-4264-1258}
\affiliation{%
  \department{School of Medicine}
  \institution{University of California San Diego}
  \city{La Jolla}
  \state{CA}
  \country{United States}
}
\email{mihogarth@ucsd.edu}

\author{Alison A. Moore}
\orcid{0000-0003-2989-4346}
\affiliation{%
  \department{School of Medicine}
  \institution{University of California San Diego}
  \city{La Jolla}
  \state{CA}
  \country{United States}
}
\email{alm123@ucsd.edu}

\author{Emilia Farcas}
\orcid{0000-0001-6485-0141}
\affiliation{%
  \department{Qualcomm Institute}
  \institution{University of California San Diego}
  \city{La Jolla}
  \state{CA}
  \country{United States}
}
\email{efarcas@ucsd.edu}

\author{Nadir Weibel}
\orcid{0000-0002-3457-4227}
\affiliation{%
  \department{Computer Science and Engineering}
  \institution{University of California San Diego}
  \city{La Jolla}
  \state{CA}
  \country{United States}
}
\email{weibel@ucsd.edu}

\renewcommand{\shortauthors}{Chen~\etal}

\begin{abstract}
While voice user interfaces offer increased accessibility due to hands-free and eyes-free interactions, older adults often have challenges such as constructing structured requests and perceiving how such devices operate.  
Voice-first user interfaces have the potential to address these challenges by enabling multimodal interactions. Standalone voice $+$ touchscreen Voice Assistants~(VAs), such as Echo Show, are specific types of devices that adopt such interfaces and are gaining popularity. 
However, the affordances of the additional touchscreen for older adults are unknown. 
Through a $40$-day real-world deployment with older adults living independently, we present a within-subjects study (N = $16$; age $M = 82.5$, $SD = 7.77$, $min. = 70$, $max. = 97$) to understand how a built-in touchscreen might benefit older adults during device setup, conducting self-report diary survey, and general uses.
We found that while participants appreciated the visual outputs, they still preferred to respond via speech instead of touch.
We identified six design implications that can inform future innovations of senior-friendly VAs for managing healthcare and improving quality of life.
\end{abstract}

\maketitle

\section{introduction}\label{sec::intro}
Voice user interfaces offer increased accessibility due to the nature of hands-free and eyes-free interactions~\cite{Pradhan2018, Sunyoung2021Exploring, Hanley2021}. However, older adults often have challenges such as constructing structured requests and perceiving how such devices operate~\cite{Kim2021}.  
Touchscreen-based voice-first user interfaces, referring to those that {\it ``primarily accept user input via voice commands, and may augment audio output with a tightly integrated screen display''}~\cite{Whitenton2017, Thompson2008}, have the potential to address the aforementioned challenges by enabling multimodal interactions~\cite{Goebel2020}.
Standalone voice $+$ touchscreen Voice Assistants~(VAs) (\eg~Echo Show~\cite{echo_show}) are specific types of devices that adopt such interfaces and are gaining popularity among young people~\cite{Goebel2020}. 
However, the affordances~\cite{Norman2013} of the additional touchscreen --- both in terms of helping users and preventing them from achieving their goals --- are still unknown for older adults.
For example, it is unclear if and how older adults appreciate the affordances of the touchscreen of a device like the Echo Show~\cite{echo_show} \mbox{(Fig.~\ref{fig::teaser}e - j)} in comparison to its voice-only counterpart (\ie~Echo Dot~\cite{echo_dot}, Fig.~\ref{fig::teaser}a - d); the Echo Show allows users to input commands through {\it touch} or speak, and {\it see} additional visual elements along with the audio response, yet it also brings setbacks such as larger form-factor and more complicated visual interfaces.

Existing research has explored how older adults have used and perceived existing features of voice-{\it only} VAs \mbox{\cite{Chen2021assets, Milka2020, Pradhan2019, Pradhan2020, Sunyoung2021}}, but the potential merits and setbacks of the secondary touchscreen modality are underexplored.
Additionally, most prior works~(\eg~\cite{Milka2020, Pradhan2019}) only focused on general uses~(\eg~music) of VAs for older adults, rather than healthcare applications, which are indispensable use cases~\cite{Chen2021assets, Charles2021}.

We present a real-world within-subjects study~(N = $16$; age $M = 82.5$, $SD = 7.77$, $min. = 70$, $max. = 97$) to understand the affordances that the additional touchscreen for standalone VAs could bring to aging populations.
Besides general uses~\mbox{(\eg~\cite{Milka2020, Sunyoung2021})}, we investigate the feasibility of using voice to collect self-report End-of-Day~(EOD) diary survey (hereafter referred to as {\it diary})~\cite{EMA}, which help clinicians better understand older adults' life routines and healthcare needs.
Our study, based on Echo Dot and Show, aims to address three key Research Questions (RQs):

\begin{itemize}[leftmargin=*]

  \item \textbf{(RQ1)} How does the built-in touchscreen affect the older adults' experience of {\it setting up the device}?

  \item \textbf{(RQ2)} How does the built-in touchscreen affect the older adults' behaviors and experience of {\it conducting self-report daily diary survey}?

  \item \textbf{(RQ3)} How does the built-in touchscreen affect the older adults' behaviors and experience of using VAs for {\it general purposes}?
\end{itemize}

In collaboration with UC San Diego Health and the Vi at La Jolla, we deployed both devices in real-world older adults' residences for a total of $40$~days.
Our findings include three aspects:
\textbf{(1)} During device setup, older adults appreciated the merits of the additional touchscreen. 
Quantitatively, we measured an overall reduction in the task completion time of around $50$\% while setting up the Echo Show, compared to the counterpart;
\textbf{(2)} While conducting self-report diary survey, participants suggested that both systems needed to be more interactive and conversational. But overall, they enjoyed the visual output enabled by the additional display. Despite this, participants still responded more often to survey questions via speech over touch, although the time needed for touch input was characterized by an approximately $20$\% reduction;
\textbf{(3)} For general uses, older adults appreciated the visual outputs and acknowledged the sense of companionship with the voice-first modality. The additional visual information (\eg~visual texts and icons) also encouraged older adults to engage more with the VAs. However, also in this case, interactions were mostly based on speech input instead of touch.
Based on these findings, we identified six design implications under two themes that can inform future innovations of senior-friendly VAs for managing healthcare and improving Quality of Life~(QOL).
We believe our work will impact practitioners and researchers attempting to design senior-friendly voice-first VAs for aging populations, both for enhancing their healthcare and to better support general uses.

\section{Related Work}\label{sec::related}
\subsection{Design of Voice-Based Virtual Assistants~(VAs)}\label{sec::related::va}
\begin{figure}
    \includegraphics[width=0.32\textwidth]{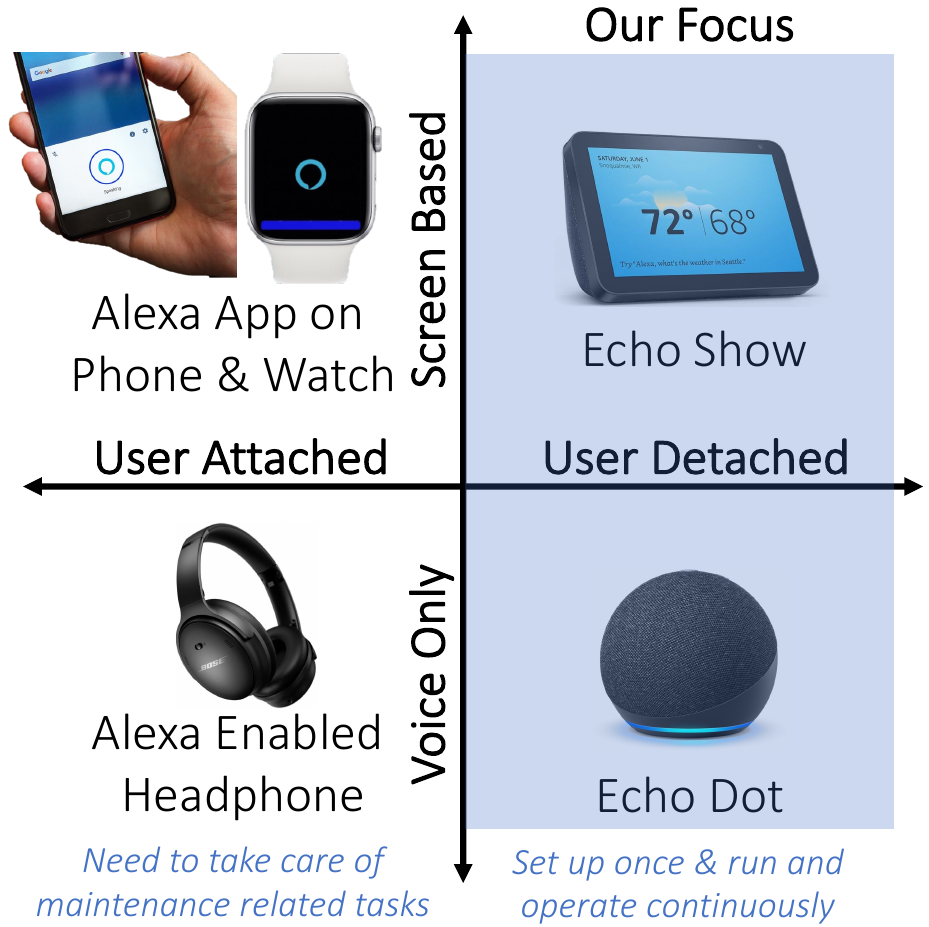}
    \Description{Design taxonomy of voice-enabled smart devices. The \textit{user-attached}, \textit{multi-modal} design would provide the highest interactively, yet could potentially introduce the complexities for older adults during operation, setup, and debugging phases.}
    \vspace{-0.1in}
    \caption{Design taxonomy of VAs. On one extreme, the \emph{user-attached} and \emph{voice $+$ touchscreen} design provides the highest interactivity, yet might introduce complexity during operation, setup, and troubleshooting. On the other extreme, the \emph{user-detached} (or \emph{standalone}) and \emph{voice-only} design offers the simplest way for device uses, yet provides the lowest interactivity.}
    \Description{Figure 2 describes a taxonomy diagram showing the design paradigm. The Alexa App on a phone and watch is considered as user-attached and touchscreen-based. The Echo Show is considered as user-detached and touchscreen-based. The headphone with built-in Alexa is considered as user-attached and voice-only. The Echo Dot is considered to be user-detached and voice-only.}
    \vspace{-0.15in}
    \label{fig:design_taxonomy}
\end{figure}

VAs refer to software agents that listen and respond to verbal commands~\cite{Coates2019} and have been integrated into a \mbox{heterogeneous} hardware embodiment.
Fig.~\ref{fig:design_taxonomy} shows a taxonomy based on two dimensions: \textit{modality} (\ie~voice-only or voice $+$ touchscreen) and \textit{the way the devices are incorporated into users' lives} (\ie user-attached or -detached).

\vspace{4px}
\noindent{\bf Voice \vs~ Voice $+$ Touchscreen.}
The simplest VA embodiment is the smart speaker, where voice is the {\it only} \mbox{supported} modality for both queries and responses. 
While offering hands-free and eyes-free control, voice-only interfaces present two major issues.
First, receiving information only by voice is often ambiguous and inefficient for \mbox{information} \mbox{consumption} due to sequential information access, in contrast to the visual scanning~\cite{Whitenton2017}.
Second, \mbox{voice-based} \mbox{interactions} yield more turn-taking compared to screen-based interactions~\cite{Reicherts2022}.
To address these, existing research \mbox{proposed} the concept of touchscreen-based voice-first user interfaces, referring to those VAs whose primary \mbox{functionality} can be accessed through speech, but have a touchscreen as an auxiliary medium for information input and \mbox{output} \mbox{\cite{Thompson2008, Bajorek2018}}.
Such interfaces bring together the merits of voice (an efficient input modality) and screen (an efficient \mbox{output} \mbox{modality})~\cite{Whitenton2017}.
Furthermore, recent works (\eg~\cite{Bravo2020, Mehrotra2016}) demonstrated that the embodied conversational agents --- VAs that leverage the display to show a visual representation of a human~\cite{Bickmore1999} --- are preferred by older adults in terms of social isolation and loneliness.

\vspace{4px}
\noindent{\bf User-Attached \vs User-Detached devices.}
User-attached VAs are voice-enabled devices that need to be worn or handheld by users (\eg Siri on iPhone), and are involved with repetitive maintenance related tasks (\eg~charging devices).
In contrast, standalone VAs (\eg~smart speakers and displays) usually need an external power supply and are expected to be placed in a particular environment permanently.
These devices are promising for older adults, as they only need to be set up once and can run continuously, so users can focus on task-related operations (\eg~making queries)~\cite{Chen2021assets}. 
Our study only involved user-detached (\ie~standalone) devices.

\subsection{Usability Evaluation and Real-World Deployment of Voice User Interfaces for Older Adults}\label{sec::related::older_adult_iva}
While existing research recognized the promise of using voice for information interactions for older adults~\cite{Chen2021assets, Stigall2019}, most work focused on evaluating voice-{\it only} VAs from the older adults' perspective.
For example, Jes{\'u}s-Azabal \mbox{\etal} \cite{Azabal2019} introduced {\it Remembranza}, a medication reminder skill based on Echo speakers.
Trajkova~\etal~\cite{Milka2020} investigated older adults' uses of smart speakers, and found that most participants become non-users due to the lack of perceived usefulness.
Upadhyay~\etal~\cite{Upadhyay2023} studied the explorations and long-term uses of VAs by older adults living in a long-term care community, however the affordances of touchscreen were not explored. 
Kakera~\etal~\cite{Karkera2023} identified the usefulness of multiple features that can be realized by VAs for supporting older adults living independently, yet the voice-first VAs were excluded.
Choi~\etal~\cite{Choi2018} suggested the most frequently used features of Echo speaker by older adults are asking practical questions and managing tasks.
Few recent works \cite{Milka2020, Chen2021, Jansons2022} emphasized that although voice-only VAs do not require older adults to be technology savvy, the low interactivity might fail to support older adults' needs, especially when it comes to the management of health data.
Similarly, \mbox{Nallam~\etal~\cite{Nallam2020}} suggested that older adults see the potentials for using VAs to search for health information and support health tasks, yet adoptions of VAs could be affected by access barriers, confidentiality risks, and receiving trusted information. 
\mbox{Pradhan~\etal~\cite{Pradhan2019}} investigated how older adults treat the smart speakers as a human.
They also explored how Echo Dot could be used by older adults with low technology experience~\cite{Pradhan2020}.
\mbox{Bonilla \etal~\cite{Bonilla2020}} explored older adults' understanding of VAs' privacy and security implications.
\mbox{Kim~\etal~\cite{Sunyoung2021}} conducted a longitudinal study to understand older adults' perception and use of Google Mini~\cite{google_mini}.
They also explored the initial interactions of older adults while using a smart speaker.
\mbox{Shade~\etal~\cite{Shade2021}} focused on medication reminders using Google Home Mini.

Researchers also investigated heterogeneous multi-modal voice-first interfaces to help improve older adults' QOL.
\mbox{Shalini~\etal~\cite{Shalini2019}} designed a system for older adults to track health information~(\eg sleep quality) inferred by \mbox{instrumented} in-home sensors, using both audio and visual display.
\mbox{Barros~\etal~\cite{Barros2020}} conducted a usability \mbox{assessment} of smartphone Google Assistant and Siri, and found that users prefer the Siri interface because it is minimalist.
\mbox{Hu~\etal~\cite{Hu2022}} showed the designs of seven types of speech acts for older adults by leveraging the built-in touchscreen using \mbox{politeness} theory.
\mbox{Gustafson~\etal~\cite{Gustafson2022}} showed that using touchscreen-based VAs for delivering eHealth \mbox{interventions} is more \mbox{effective} compared to using laptops.
Further, researchers also investigated the integration of VAs with \mbox{heterogeneous} smart home devices.
For example, \mbox{Kowalski~\etal~\cite{Kowalski2019}} studied how VAs benefit older adults when they are integrated with smart home technologies. 
\mbox{Ennis~\etal~\cite{Ennis2017}} incorporated Echo into smart cabinets to support older adults' \mbox{independence}.
\mbox{Valera Rom{\'a}n~\etal~\cite{Roman2021}} studied how the combination of smart bracelets, smart home devices, and VAs could allow older adults to monitor their physical activity and sedentary patterns.

In contrast to existing research that only focuses on voice-only VAs and/or the techniques to design touchscreen-based VAs for specific types of interactions, we explored {\it how the voice $+$ touchscreen VAs could influence older adults' experience of device setups, conducting self-report diaries, and general uses} through a real-world deployment.

\begin{figure*}[t]
    \centering
    \includegraphics[width=\textwidth]{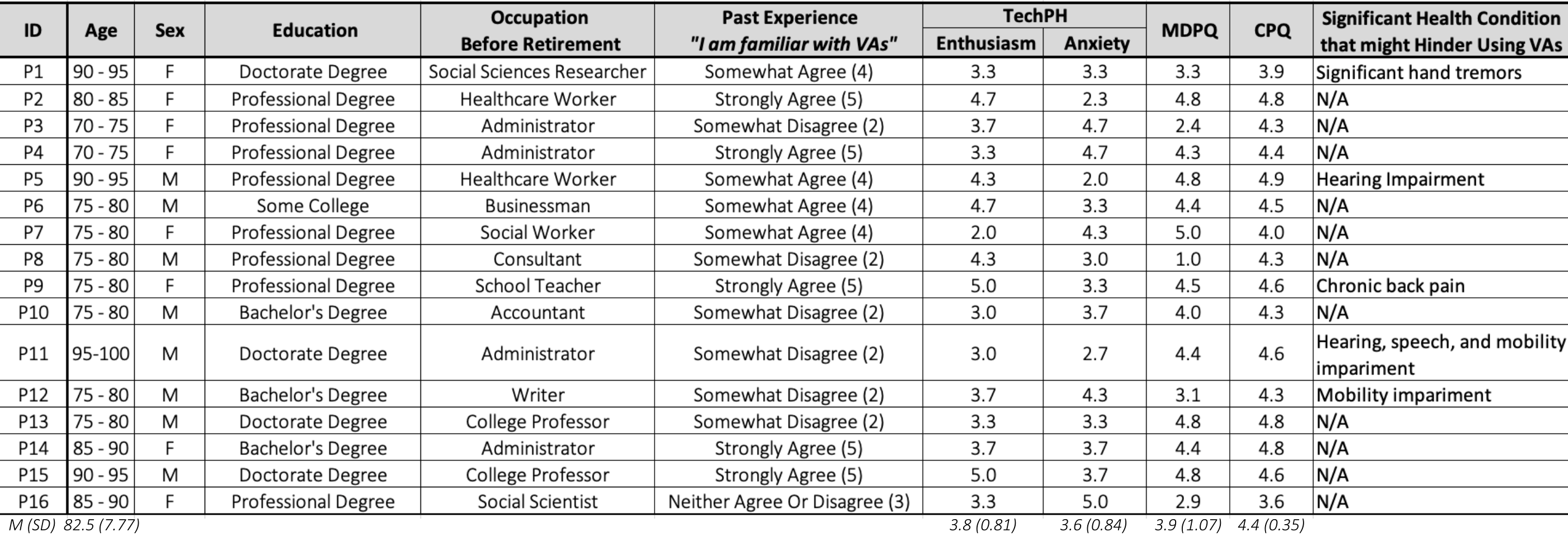}
    \vspace{-0.30in}
    \caption{Participants' demographics. All self-reported scores are on the scale of 1 to 5. Participants proficient with technology would have high score of \emph{VA Past Experience}, \emph{TechPH (Enthusiasm)}, \emph{MDPQ}, and \emph{CPQ}, and a low score of \emph{TechPH (Anxiety)}.}
    \Description{Figure 3 shows participants' demographic data. P1 is in the range of 90 to 95 years old. She has a doctorate degree and was a social sciences research before retirement. She rated the past experience, enthusiasm, anxiety, MDPQ and CPQ as 4, 3.3, 3.3, 3.3 and 3.9. P2 is in the range of 80 to 85 years old. She has a professional degree and was a healthcare worker before retirement. She rated the past experience, enthusiasm, anxiety, MDPQ and CPQ as 5, 4.7, 2.3, 4.8 and 4.8. P3 is in the range of 70 to 75 years old. She has a professional degree and was an administrator before retirement. She rated the past experience, enthusiasm, anxiety, MDPQ and CPQ as 2, 3.7, 4.7, 2.4 and 4.3. P4 is in the range of 70 to 75 years old. She has a professional degree and was an administrator before retirement. She rated the past experience, enthusiasm, anxiety, MDPQ and CPQ as 5, 3.3, 4.7, 4.3 and 4.4. P5 is in the range of 90 to 95 years old. He has a professional degree and was a healthcare worker before retirement. He rated the past experience, enthusiasm, anxiety, MDPQ and CPQ as 5, 5.3, 2.0, 4.8 and 4.9. P6 is in the range of 75 to 80 years old. He has some college experience and was a businessman before retirement. He rated the past experience, enthusiasm, anxiety, MDPQ and CPQ as 4, 4.7, 3.3, 4.4 and 4.5. P7 is in the range of 75 to 80 years old. She has a professional degree and was a social worker before retirement. She rated the past experience, enthusiasm, anxiety, MDPQ and CPQ as 4, 2.0, 4.3, 5.0, and 4.0. P8 is in the range of 75 to 80 years old. He has a professional degree and was a consultant before retirement. He rated the past experience, enthusiasm, anxiety, MDPQ and CPQ as 2, 4.3, 3.0, 1.0, and 4.3. P9 is in the range of 75 to 80 years old. She has a professional degree and was a school teacher before retirement. She rated the past experience, enthusiasm, anxiety, MDPQ and CPQ as 5, 5, 3.3, 4.5, and 4.6. P10 is in the range of 75 to 80 years old. He has a bachelor’s degree and was an accountant before retirement. He rated the past experience, enthusiasm, anxiety, MDPQ and CPQ as 2, 3.0, 3.7, 4.0, and 4.3. P11 is in the range of 95 to 100 years old. He has a doctorate degree and was an administrator before retirement. He rated the past experience, enthusiasm, anxiety, MDPQ and CPQ as 2, 3.0, 2.7, 4.4, and 4.6. P12 is in the range of 75 to 80 years old. He has a bachelor’s degree and was a writer before retirement. He rated the past experience, enthusiasm, anxiety, MDPQ and CPQ as 2, 3.7, 4.3, 3.1, and 4.3. P13 is in the range of 75 to 80 years old. He has a doctorate degree and was a college professor before retirement. He rated the past experience, enthusiasm, anxiety, MDPQ and CPQ as 2, 3.3, 3.3, 4.8, and 4.8. P14 is in the range of 85 to 90 years old. She has a bachelor’s degree and was an administrator before retirement. She rated the past experience, enthusiasm, anxiety, MDPQ and CPQ as 5, 3.7, 3.7, 4.4, and 4.8. P15 is in the range of 90 to 95 years old. He has a doctorate degree and was a college professor before retirement. He rated the past experience, enthusiasm, anxiety, MDPQ and CPQ as 5, 5.0, 3.7, 4.8, and 4.6. P16 is in the range of 85 to 90 years old. She has a professional degree and was a social scientist before retirement. He rated the past experience, enthusiasm, anxiety, MDPQ and CPQ as 3, 3.3, 5.0, 2.9, and 3.6.}
    \vspace{-0.15in}
    \label{fig::demographic}
\end{figure*}

\subsection{Ecological Momentary Assessments (EMA) and End-Of-Day (EOD) Diaries Data Collections}\label{sec::related::ema}
Ecological Momentary Assessment~(EMA) involves repeated sampling of subjects' current behaviors and experiences in real time in their environment~\cite{Shiffman2008}.
EMA can assist medical providers to better understand patients' daily routines and healthcare needs, particularly important for older adults with chronic diseases~\cite{Chen2021assets}.
Moskowitz~\etal~\cite{Moskowitz2006} classified EMA into three types: {\it diaries} (fixed interval assessment with a frequency of once per day, employing retrospective coverage strategy~\cite{Shiffman2008}), {\it experience sampling} (using specific signaling devices that randomly notify participant to make reports a fixed number of times per day~\cite{Moskowitz2006}), and {\it event-based sampling} (self-reports are solicited at the time the variable of interest such as physical activity takes place~\cite{Bolger2003}).

While online survey is a common and simple strategy for collecting EMA data (\eg~~\cite{Compton2008}), this method is usually limited in terms of compliance and accessibility.
Some researchers also explored using smartphones to collect diaries (\eg~\cite{Aminikhanghahi2020}) possibly because of the widely accessible of smartphones compared to computers.
However, the low efficiency of typing due to finger dexterity problems~\cite{Vogel2007} and the complexities of troubleshooting these devices might be problematic for older adults.
The recent $\mu$EMA~\cite{Intille2016} demonstrated the effectiveness of using smartwatches to conduct microinteraction-based event sampling, through which participants could answer questions by a quick tap on the smartwatch.
However, typing on a smartwatch for open-ended questions is impractical.

Voice has been identified as a promising approach for collecting survey data from older adults~\cite{Chen2021assets, Charles2021, Lifset2020}.
Prior researchers have used interactive voice response systems to address the limitations of mobile and wearable devices~\cite{Courvoisier2010}. 
Instead, we focus on using {\it standalone} VAs for {\it self-report diary} data collection, where participants need to retrospectively respond to a set of questions for the past $24$ hours.
We focus on standalone devices, for two reasons.
First, unlike the young adults, older adults are more likely to be at home more often~\cite{Cabrita2017, Mannell1991}.
This phenomenon was more prominent during the COVID-19 pandemic and the social distancing restrictions, particularly for older adults who are known to be a group at higher risk~\cite{cdc_older}.
Second, we only focus on {\it diary} studies that are not time sensitive~(Sec.~\ref{sec::method::ema} and Appendix~\ref{sec::app::ema_question}).
Participants were expected to complete the survey on a daily basis, but the time did not have to be strictly specified.

\begin{figure*}[t]
    \centering
    \includegraphics[width=\textwidth]{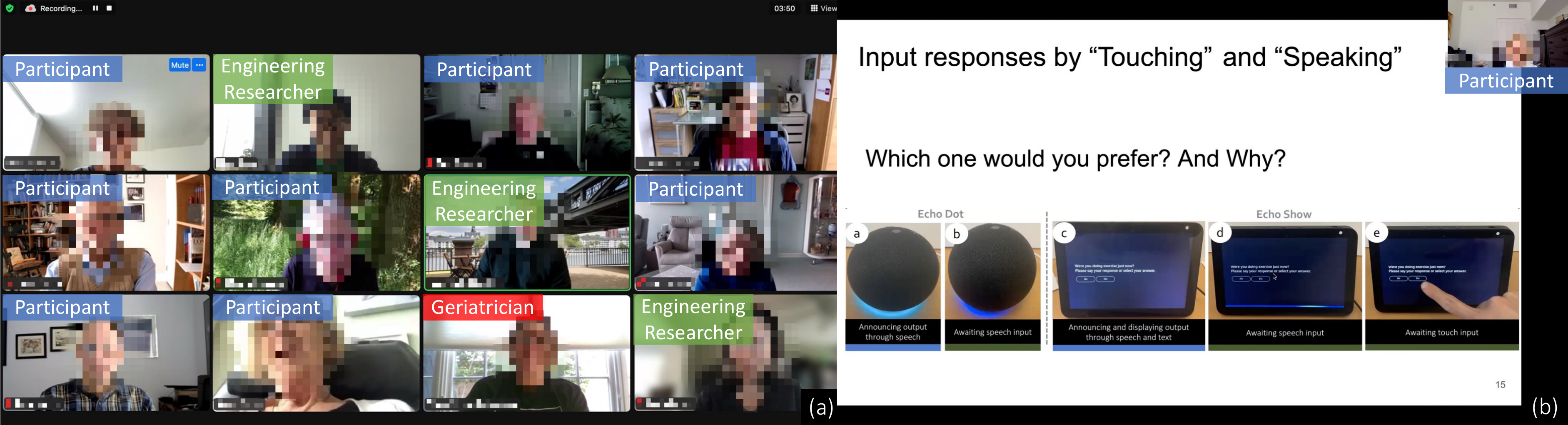}
    \vspace{-0.25in}
    \caption{(a) One of the focus groups conducted through Zoom with eight participants, three engineering researchers, and one geriatrician; (b) One participant (top right corner) was explaining her ideas of Q6 of Fig.~\ref{fig::app::guiding_questions} in Appendix~\ref{sec::app::guiding_questions}.}
    \Description{Figure 4a shows the screenshot of the Zoom session of one remote focus group session. Eight participants, three engineering researchers and one geriatrician were presented and moderate the discussions. Figure 4b describes the screenshot when one older adult participant was explaining and justifying her responses, regarding the prompt of ``which input method would you prefer? Touching or speaking?'' Demonstrative figures were shown to help participants better recall their experience;}
    \vspace{-0.1in}
    \label{fig::focus_group}
\end{figure*}

\section{Methods}\label{sec::method}
\subsection{Participants and Study Procedures}\label{sec::method::design}
We recruited $16$~older adults, including eight males and eight females, through UC San Diego Health\footnote{UC San Diego Health: \url{https://health.ucsd.edu} [Accessed on 7/1/2023]} and the Vi at La Jolla Village\footnote{The Vi at La Jolla Village: \url{https://www.viliving.com/locations/ca/san-diego-la-jolla} [Accessed on 7/1/2023]} (age, $M = 82.5$, $SD = 7.77$, {\it min.} = $70$, {\it max.} = $97$, see Fig.~\ref{fig::demographic}).
All participants are self-identified as capable of living independently.
The study was approved by the Institutional Review Board~(IRB), and the Echo Dot and Echo Show ($\sim$\$$130$) were provided as incentives.
Our study was conducted during December to February, during which all participants resided at home during majority of the time without travel plans.
Overall, our study was structured in four phases, lasting for $40$~days at the residences of older adults:

\vspace{4px}
\noindent{\bf Phase 1: Pre-Study Questionnaires.}~Participants completed informed consent and questionnaires, including their past experience using VAs, as well as three validated questionnaires investigating their attitudes towards technology and their experiences with it (see Fig.~\ref{fig::demographic}). 

\vspace{4px}
\noindent$\bullet$ {\bf TechPH~(Technophilia)}~\cite{Anderberg2019} measures the older adults' attitudes toward technology. We reported the average score for their {\it enthusiasm} and {\it anxiety} toward general technologies. 
A high enthusiasm and low anxiety imply a positive attitude as measured through this instrument.

\vspace{4px}
\noindent$\bullet$ {\bf MDPQ~(Mobile Device Proficiency Questionnaire)}~\cite{Roque2018, Petrovcic2019} evaluates the proficiency with \mbox{smartphones} for older adults.
While smartphones are not our focus, we extrapolated that the older adults would transfer existing skills from \mbox{interactions} with smartphones to voice-based interactions.
A high MDPQ score indicates a highly proficient smartphone user.

\vspace{4px}
\noindent$\bullet$ {\bf CPQ~(Computer Proficiency Questionnaire)}~\cite{Boot2015} evaluates the older adults' proficiency with desktop computers.
Similar to the MDPQ, we hypothesized that older adults might transfer some existing computer skills to voice-based interactions.
A high CPQ score indicates a highly proficient desktop computer user.

Notably, while focusing on VAs, the results of MDPQ and CPQ could reflect and imply older adults’ skills, experience, and attitude toward using VAs from a broader context by looking at richer perspectives of technology exposures and uses in their life.

\vspace{4px}
\noindent{\bf Phase 2: Device Setup.}
Our second phase of the study aims to address RQ1.
Although VAs may often be set up by others, it is still important for older adults to know how to initialize the device and finish the last-mile tasks~(\eg~connect the device to the WiFi)~\cite{Chen2021assets, Chen2023VOLISetup}.
A {\it within-subject design} was used to evaluate participants' performance on setting up VAs with and without touchscreen.
We pre-setup and initialized the devices with our Alexa skill, yet we required participants to engage in the last-mile tasks. 
An experimental Amazon account was created for each participant, enabling us to track participants' responses and usages.
During in-person meetings, we first introduced the scope of our project, described and had participants signed all required forms, and explained the technology related to VAs.
We then invited participants to set up two devices {\it independently}.
Participants were asked to use the official instruction manuals as needed.
Eight participants were prompted to setup Echo Dot first, followed by Echo Show. 
While another eight participants were prompted to setup Echo Show first, followed by Echo Dot.
While setting up the devices, participants were required to type the username and password of our pre-created experimental account, with the average number of characters being $33$ ($SD = 2.28$) mixed with letters, numbers and special characters.
Assistance was provided, {\it if and only if} the participants gave up on the effort.
Semi-structured interviews were conducted after participants set up each VA.
The questions used to guide the interviews could be referred to Appendix~\ref{sec::app::guiding_questions}.
While we excluded the quantitative measures from participants who gave up on setting up device(s), participants were still encouraged to discuss about their feeling after observing the research assistant(s) setting up the device(s) on their behalf.
This phase on average spent $37.03$~min ($SD = 9.50$~min) with each participant.

\vspace{4px}
\noindent{\bf Phase 3: Investigations of VA Uses for Conducting Self-Report Diaries and General Uses.}~
We used a {\it within-subject design} to evaluate the values of the additional touchscreen while using VA to conduct self-report diary survey (RQ2) and for general uses (RQ3).
Eight participants were asked to use Echo Dot first, followed by Echo Show, each for $15$~days, while another eight participants were instructed to use Echo Show first, followed by Echo Dot.
Before each $15$-day session, each participant was instructed on how to use VAs with official user manuals and provided with five days to explore and familiarize themselves with the given devices.
During each $15$-day session, participants were instructed to use the features of their VAs for general uses for added convenience and benefit to their daily routine.
Additionally, since we aimed to explore the feasibility and usability of using VAs as a tool for conducting self-report daily diaries, we designed a set of diary questions for wellness screening~(Sec.~\ref{sec::method::ema}). 
Participants were expected to complete the diary on a daily basis, but were not required to complete it at specific time of the day, and could choose when to engage with the device. 
At the end of the diary survey, a usability question was delivered to each participant, using the prompt: {\it ``on a scale of 1 – 5, how do you like to use voice assistant to report your diary survey? 1 being dislike extremely and 5 being like extremely''}.
This offered us insights on the participants' overall experience after each time they used VAs for the diary survey.
Participants were encouraged to choose their preferred methods to remind themselves.
At the end of each $15$-day session, participants were invited to rate how strongly they agreed with the following three statements in a $5$-point Likert scale:

\begin{itemize}[leftmargin=*]

  \item \textbf{(Q1)}~{\it ``Conducting the daily diary using the given VA could cause interruption burden to my daily life routine''};
  
  \item \textbf{(Q2)}~{\it ``I am comfortable to use the VA for general uses''};
  
  \item \textbf{(Q3)}~{\it ``It is easy to use the VA for general uses''};

\end{itemize}

Participants were then invited to complete NASA TLX~\cite{nasatlx} and System Usability Scale~(SUS)~\cite{Brooke1996, Brooke2013} questionnaires regarding their overall user experience.
To minimize the time and effort needed to complete the questionnaires, we excluded the pair-wise workloads comparisons in TLX, and assumed the weights for each perceived workloads were identical while computing the overall TLX score. 
We then conducted a remote semi-structured interview with each participant. All procedures were repeated for the second $15$-day session with the other device.

\vspace{4px}\noindent{\bf Phase 4: Focus Groups.}~
We adopted Robson~\etal's suggestion~\cite{Robson2002} regarding the size of the focus group to be eight to twelve for an in-depth discussion, and therefore organized two online focus groups (Fig.~\ref{fig::focus_group}). 
The attendees of each focus group included eight participants, three engineering researchers, and one geriatrician. 
Same prompts (Fig.~\ref{fig::app::guiding_questions} in Appendix~\ref{sec::app::guiding_questions}) and slides were used to guide the discussion in both focus groups.

\subsection{Design of the Diary Survey}\label{sec::method::ema}
To understand the affordances of the touchscreen while using standalone VAs to collect diaries, we designed a self-report diary survey with geriatricians from \emph{the Anonymous Academic Medical Center} and empirical guidance from the World Health Organization~\cite{whoaging}. 
Our diary survey was centered around the eight themes: quality of sleep, social interactions, exercise, pain management, alcohol uses, food consumption, symptoms, and medication management. 
We also added a set of usability questions for the purpose of our study (Appendix~\ref{sec::app::ema_question}).
%
% content is not important as we only evaluate feasibility.
While establishing the validity of diary data acquired is beyond our scope, we have iteratively designed questions with one geriatrician that could be answered easily and quickly (the whole survey would typically take $\leq$ $5$~min and could be interrupted at any time, verified during pilot testing).

% type of questions, and structures of the answer
We expected four types of answers: {\it binary}, {\it Likert}, {\it number}, and {\it open}.
The design of the first three types of responses has been widely used in many existing EMA studies~(\eg~\cite{Dunton2012, Intille2016}). 
Participants were asked to speak their choice, and were provided with the additional alternative option to input response by {\it touching} the buttons on the touchscreen while using the Echo Show (Fig.~\ref{fig::teaser}i).
%
% why only 5 buttons
We adopted the same design as $\mu$EMA~\cite{Intille2016}, and provided five options (\ie~five buttons) for {\it Likert}- and {\it number}-type responses with touch input.
We designed the height of each button to be approximately the same as the finger width, and the width of the button being around twice the finger width (Fig.~\ref{fig::teaser}i).
This decision was made to ensure all buttons are easy to be clicked while fitting on the same screen.
Following the suggestion from the geriatrician, we also included an {\it open} unstructured type of response that supported only the input modality via speech.
We did not include healthcare feedback to the participants' reported data due to liability concerns from our institution's IRB. 
Our study only focused on data collection and observation of behavior, not intervention to modify behavior.

\subsection{Implementation}\label{sec::methods::implementations}
We selected Echo Dot~\cite{echo_dot} (Fig.~\ref{fig::teaser}a - d) and Echo Show with a built-in eight inches touchscreen~\cite{echo_show} (Fig.~\ref{fig::teaser}e - j) as the testbed due to the dominant market share~\cite{Kinsella2020}.
However, the majority of our findings could be transferred to other similar standalone VAs.
We implemented an Alexa skill and a Flask backend to track the conversation states.
{\it``My Health''} was used as the invocation name, which could be easily remembered and clearly spoken as verified during our pilot testing.
\subsection{Measures and Data Analysis}\label{sec::method::measures}
We structured our analysis based on the experience of device setup \textbf{(RQ1)}, conducting self-report daily diaries \textbf{(RQ2)}, and general uses \textbf{(RQ3)}.
We describe the measures and approaches of analyzing data for each aspects.

\begin{figure*}[t]
    \vspace{-0.15in}
    \centering
    \includegraphics[width=\textwidth]{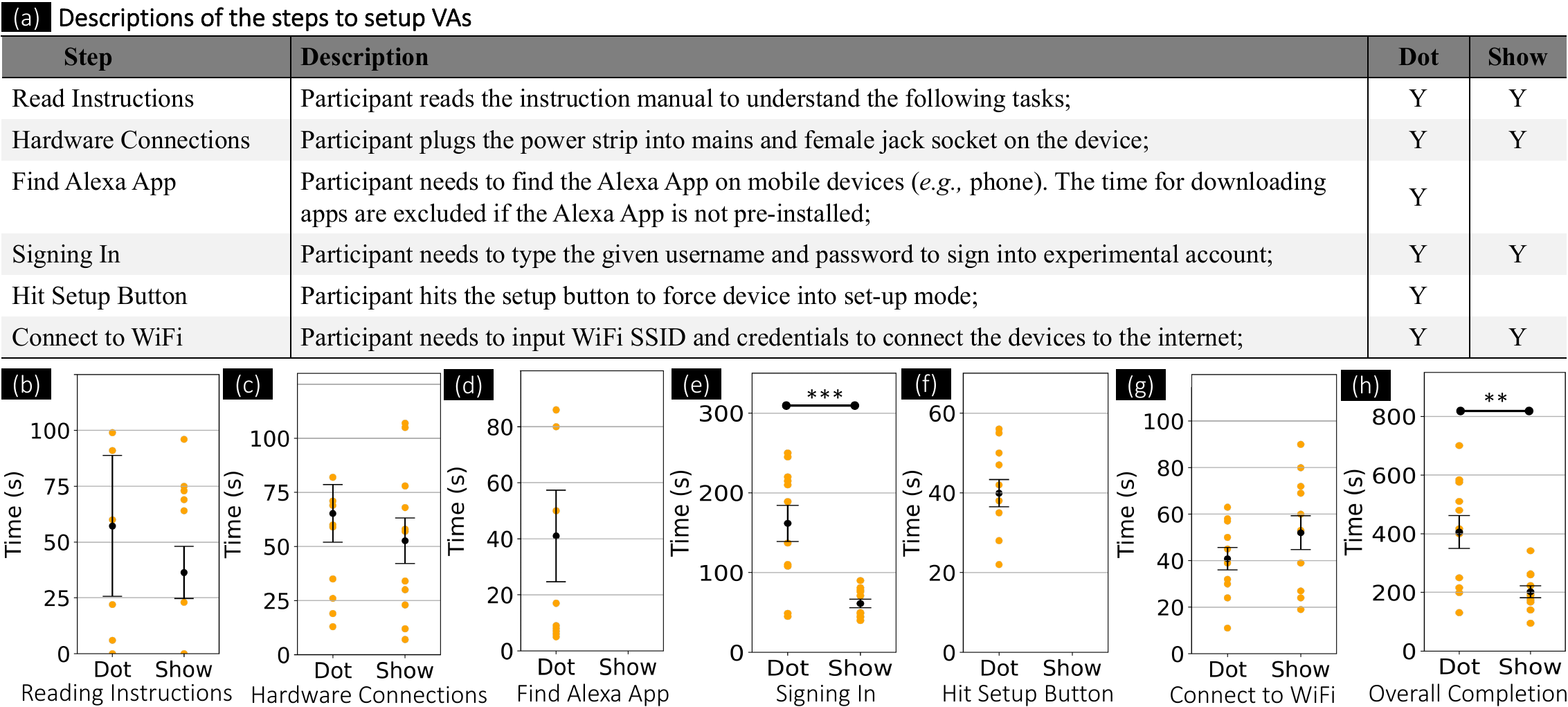}
    \vspace{-0.25in}
    \caption{Evaluation results for participants setting up Echo Dot and Echo Show. (a) Description of steps for setting up VAs; (b) Task Completion Time~(TCT) for reading instructions, (c) hardware connections, (d) finding the Alexa mobile app, (e) signing in with the experimental account, (f) finding and hitting the setup button, (g) connecting to WiFi, and (h) the overall TCT while setting up Echo Dot and Show. Notably, setting up Echo Show does not require participants to use the mobile Alexa app (d) and press the setup button (f). Participants (P4, P6, P7, P11, P12) who did not complete the tasks or gave up on the whole session were not included. We used standard error to represent the error bar. Notations for indicating the statistical significance of \posthoc~test: * = $ .05 > p \ge .01$, ** = $ .01 > p \ge .001$, *** = $p < .001$.}
    \Description{Figure 5a shows the workflow to set up Echo Dot and Echo Show. Setting up the Echo Dot needs to follow the steps of read instruction, hardware connections, find Alexa App on the mobile devices, sign in, hit the setup button, and connect the WiFi. Setting up the Echo Show needs to follow the steps of read instruction, hardware connections, sign in, and connect to WiFi. Figure 5b shows the task completion time for reading instructions while setting up the Echo Dot and Echo Show. No statistical significance is found. Figure 5c shows the task completion time for hardware connections while setting up the Echo Dot and Echo Show. No statistical significance is found. Figure 5d shows the task completion time for finding Alexa Apps on the mobile phone/tablet while setting up the Echo Dot. Figure 5e demonstrates the task completion time for signing into the experimental account while setting up the Echo Dot and Echo Show. Statistical significance was found with ``***'' being annotated. Figure 5f describes the task completion time for forcing Echo Dot to enter the setup mode. Figure 5g shows the task completion time for connecting device to the WiFi while setting up the Echo Dot and the Echo Show. No statistical significance is found. Figure 5h demonstrates the overall task completion time for setting up the Echo Dot and Echo Show. Statistical significance was found with ``**'' being annotated.}
    \vspace{-0.1in}
    \label{fig::tct_setup}
\end{figure*}

\vspace{+4px}\noindent{\bf Device Setup.}
By observing participants and analyzing official instruction manuals, we first summarized six key steps (Fig.~\ref{fig::tct_setup}a) while setting up the Echo Dot and Echo Show.
While observing participants' behavior of setting up devices, we noted the Task Completion Time~(TCT) that participants spent during each step, which were then be used to compute the overall TCT.
Notably, we used the {\it accumulated} time spent on each step as the final TCT for analysis purposes, because, for example, the participant might read instructions after every setup actions interleavedly.
We did not include other actions except the six pre-defined steps in the overall TCT (such as finding the WiFi password or handling unexpected phone calls), since they are not related to the assigned tasks and vary greatly among participants. 
To analyze the data quantitatively, we used Repeated Measure Analysis of Variance~(RM-ANOVA)~($\alpha = .05$) to evaluate the statistical significance of the effects of the two devices.
The Tukey's Honestly Significant Difference~(HSD) test~\cite{Tukey1949} was used for conducting \posthoc~test.
Before performing RM-ANOVA, we first conducted the normality check of measures in each catalogue using Shapiro-Wilk test~\cite{Shapiro1965} .
For those failing to pass the normality check, we adopted Aligned Rank Transform~(ART)~\cite{art2011} for statistical significance test, followed by ART-C with Bonferroni adjustment\footnote{For the statistical analysis by ART, we reported the degree of freedoms of $F$-statistic of the aligned and ranked responses instead of original observations~\cite{art2011}. We used \texttt{ARTool} (\href{https://depts.washington.edu/acelab/proj/art/}{https://depts.washington.edu/acelab/proj/art/}) [Accessed on 7/1/2023] for conducting this analysis.}~\cite{artc2021, Bland1995}.
For all statistical significance analysis, the partial eta square ($\eta_p^2$) was used to understand the effect size, with $.01$, $.06$, and $.14$ being used as the empirical thresholds for small, medium and large effect sizes, where a larger effect size indicates a higher practical significance~\cite{Cohen2013}.
Two researchers then performed thematic analysis~\cite{Braun2006} and adopted a mixture of emergent and priori coding approaches on the $9.87$~hours of video-audio recordings.
Specifically, we first transcribed the recorded audio clips and removed the connecting phrases (\eg~{\it`` [...] you know [...]''}) to enhance readability.
We then closely read the transcripts and watched the recorded videos iteratively, and allowing codes to emerge freely from the data. 
During analysis, we held multiple discussions to discuss and iteratively refine the codes and reconcile the disagreement.
Overall, four iterations were conducted to ensure the reliability of coding results.
Fig.~\ref{fig::app::codebook_device_setups} in Appendix~\ref{sec::app::codebook} shows the codebook.

\vspace{+4px}\noindent{\bf Experience of Conducting Self-Report Daily Diaries.}
We analyzed interaction traces related to self-report diary reportings, where each {\it interaction trace} refers to a pair of request and response.
We adopted four quantitative measures that were introduced in \cite{Intille2016}, including:

\begin{itemize}[leftmargin=*]
    
    \item {\bf Diary Survey Compliance Rate}, defined by the percentage of surveys being {\it fully} completed, versus the number of surveys that were expected to be completed. A failure of survey compliance could be caused by either forgetting to start the survey, or failing to complete all designated questions by stopping mid survey. Overall, survey compliance measures the performance of the diary data collection tool on the survey level~(\ie~question set);
    
    \item {\bf Question Completion Rate}, defined by the percentage of the questions being answered over the total number of questions being delivered. Unlike survey compliance, question completion measures the feasibility of the dairy data collection tool on the individual question level;
    
    \item {\bf Initial Prompt Response Rate}, defined by the percentage of questions completed when delivered the first time. If the participant does not answer as expected (such as wrong format or without speaking any content), the system repeats the question;
    
    \item {\bf Response Latency}, measured by the elapsed time in milliseconds (ms) between the instant when a specific question is announced and the response for that particular question has been input, either by {\it voice} or {\it touch};

\end{itemize}

A similar method as device setup was then used to analyze the quantitative measures of aforementioned measures.
While performing statistical significance analysis of response latency related to touch input, we excluded participants who did not use the touch input over the course of using Echo Show.
With the same approach for analyzing qualitative data collected during device setups, we evaluated the qualitative data collected after each of $15$-day session (around a total of eight hours interviews over the phone), as well as the focus groups (around a total of two hours video-audio recordings).
Fig.~\ref{fig::app::codebook_diary_gu} in Appendix~\ref{sec::app::codebook} shows the codebook.

\vspace{+4px}\noindent{\bf Experience of Using General Features.}
With the participants' consent, we downloaded all the interaction logs captured by Amazon Privacy Portal\footnote{Amazon Privacy Portal: \href{https://www.amazon.com/alexa-privacy/apd/home}{https://www.amazon.com/alexa-privacy/apd/home} [Last accessed on 7/1/2023].} and analyzed a total of $4350$ requests related to general uses.
We manually inspected the logs and identified {\it interaction sessions} to evaluate the participants' uses of built-in features, where one interaction session contains all requests and responses when a specific skill/feature is used.
Unlike Kim~\etal~\cite{Sunyoung2021}, who used the pairs of request-response communications, interaction sessions provide a better quantitative measurements of features in use.
This is because some features (\eg~chat and knowledge query) will inherently introduce more follow-up questions compared to others (\eg~service), and therefore considering request-response pairs in isolation do not reflect usage frequencies of particular features.
Notably, the logs generated during initial five-day training sessions were not included in our analysis.
We first carefully read through all requests and responses, and used an emergent coding approach to tag each interaction session.
Similar to the qualitative analysis of interview data, three researchers analyzed the logs and discussed to refine the codes iteratively and reconcile disagreements.
To quantitative understand participants' usages, we then used the same method as device setups to evaluate the statistical significance of the frequency of the captured interaction session for each theme over the interface type, the overall NASA TLX, and SUS responses.
Semi-structured interview data collected after each $15$-day session and the focus groups related to general uses were then analyzed using the same method as device setups.
Fig.~\ref{fig::app::codebook_diary_gu} in Appendix~\ref{sec::app::codebook} shows the codebook.

\section{Results}\label{sec::result}
Our results are organized based on the three RQs, which aim to investigate the impacts of touchscreen during \textbf{(RQ1)} device setup (Sec.~\ref{sec::results::setup}), \textbf{(RQ2)} conducting self-report diary survey (Sec.~\ref{sec::results::ema}), and \textbf{(RQ3)} general uses (Sec.~\ref{sec::results::gu}).

\subsection{RQ1: How Does the Built-In Touchscreen Affect the Older Adults' Experience of Setting Up Devices?}\label{sec::results::setup}
Overall, although most participants felt it was a daunting task to set up the designated devices (\eg~{\it ``it was a little bit scary because there are a bunch of buttons and things to be pressed''}~(P13)), $11$~participants were able to fully set up both devices.
Specifically, P4 gave up during the Echo Show setup phases due to unexpected personal duties; 
P6 gave up on typing the login credentials for Dot; 
P7 gave up setting up both devices due to the lack of interest and confidence; 
P11 and P12 gave up on setting up both devices due to the inconveniences caused by impaired mobility.
Our results show that participants took significantly less TCT to set up the Echo Show than Echo Dot~($F_{1, 20} = 8.57$, $p = .003$, $\eta^2 = 0.37$, Fig.~\ref{fig::tct_setup}h).
Through qualitative analysis, we now discuss potential reasons and participants' user experience.

\begin{figure*}[t]
    \centering
    \includegraphics[width=\textwidth]{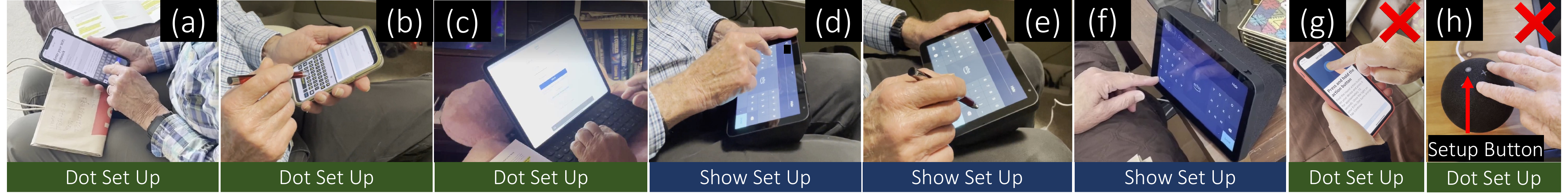}
    \vspace{-0.25in}
    \caption{Typing on Dot (a - c) and Show (d - f). (a) and (d) show the methods adopted by the majority of users. (g) shows participants incorrectly consider the button image on the phone as the setup button of the Dot. (h) shows participants incorrectly think the mute button as the setup button.}
    \Description{Figure 6a shows the demonstrative older adult was typing on a mobile phone using the finger. Figure 6b illustrates the demonstrative older adult was typing on a mobile phone using a stylus. Figure 6c shows the demonstrative older adult was typing on a tablet using an external keyboard. Figure 6d describes the demonstrative older adult was typing on the built-in touchscreen of the Echo Show using touch input while holding the Echo Show. Figure 6e shows the demonstrative older adult was typing on the built-in touchscreen of the Echo Show using a stylus while holding the Echo Show. Figure 6f describes the demonstrative older adult was typing on the built-in touchscreen of the Echo Show placed on a desktop. Figure 6g shows the demonstrative older adult incorrectly consider the button image on the phone as the setup button of the Echo Dot. Figure 6h demonstrates the demonstrative older adult correctly and successfully finds the setup button on the Echo Dot.}
    \vspace{-0.15in}
    \label{fig::echo_interactions}
\end{figure*}

\vspace{4px}
\noindent\textbf{Using the standalone touchscreen could enhance typing experience.}
Due to the lack of touchscreen of Echo Dot and the need to input credentials during the signing in phase, participants adopted different strategies to type on mobile devices, such as simply using fingers and the on-screen virtual keyboard (P3, Fig.~\ref{fig::echo_interactions}a), the stylus (P6, Fig.~\ref{fig::echo_interactions}b), or adopting a tablet with bigger display and external keyboard (P8, Fig.~\ref{fig::echo_interactions}c).
Most participants recognized the merits of the built-in touchscreen for the enhanced typing experience. 
This has been validated by our measurements where a significant reduction of TCT for the signing in stage was observed while using the Echo Show~($F_{1, 20} = 18.57$, $p < .005$, $\eta^2 = 0.48$, Fig.~\ref{fig::tct_setup}e).
In contrast, no statistical significance were observed for reading instructions ($p = .543$), hardware connections ($p = .465$), and WiFi connections ($p = .214$).
Participants' comments also reflected this observation.
For example,
P2 and P3 outlined how ease of typing is an important benefit: {\it``inputting the data is the most helpful! because the screen was bigger than my phone.''}~(P3) and {\it``the underscore sign is a little bit hard to find on this phone.''}~(P2).
P10 emphasized the merits of on-screen keyboard: {\it ``the keyboard was very different! I preferred the [the virtual keyboard on the Show], as the bigger buttons are easy to be pressed!''}~
P3 emphasized the issues that the well-known ``fat finger'' problem creates~\cite{Vogel2007}: {\it ``[with my phone] I made a lot of mistakes, because it was small. And I missed entering information with my fat fingers''}.
On the contrary, few participants still preferred to type on their phone due to the familiarity of everyday's hand-held mobile devices: 
{\it ``typing on built-in display was hard, probably because I'm more familiar with my phone''}~(P13).

\vspace{+4px}
\noindent\textbf{Immediate and \emph{in situ} visual feedback and guidance on built-in touchscreen could help track setup steps.} 
Most participants preferred the {\it visual feedback} enabled by the built-in touchscreen, to the prompts on a separate smartphone app. 
Participants highlighted the helpfulness of integrating all interaction components in one single entity, leading to a better device setting up workflow.
For example: {\it ``the touchscreen made the experience more streamlined''}~(P9).
Although P11, who needs a wheeled walker due to mobility impairment, believed that the voice based interaction should be sufficient for general uses after initial impression (\eg~{\it ``I don't need the visual, just the voice is fine''}), he still preferred the built-in touchscreen after observing the research assistant setting up both devices: 
{\it ``you don't have to worry about connecting two devices. You're dealing with one device where you have both the visual and the sound together. Whereas with the Echo Dot, you need to `plug in' a separate phone in order to get the visual!''}~
Further, many participants highlighted the benefits of immediate feedback given by the Show, and the consequent reduced demands on users' working memory~\cite{Leung2012}: 
{\it ``the setup was easier on the Show. Because we could actually see what we were doing. Whereas [with Dot] you're only hearing it and seeing it on the phone''}~(P2); 
{\it ``[Echo Show] gives me the directions right on the screen, then it would be easier than me looking at my phone and transferring the information mentally''}~(P4); 
and {\it ``[Echo Show] is better, because I have known the visual gave me immediate confirm of what I was doing''}~(P16).
In contrast, the lack of direct and \textit{in situ} visual guidance while setting up the Echo Dot could be one possible cause that led three participants to fail finding the setup button without hints.
For the Dot, part of the instructions were on the Alexa phone app. 
When being instructed to ``touch'' the setup button, some participants (\eg~P1, Fig.~\ref{fig::echo_interactions}g) incorrectly considered the button icon on the phone as the target, while others (P4, Fig.~\ref{fig::echo_interactions}h) made incorrect attempts to interact with the mute button on the Dot.
For example, P4 made this comment after being corrected by the research assistant: {\it``I had assumed the button to push was the one on the top rather than the one on the side. [...] [the system should] tell me which button to push more precisely.''}

\vspace{+4px}
\noindent\textbf{The larger physical size of the touchscreen might be a hurdle.}
Without the need of a screen, the design of the Echo Dot is naturally smaller and lighter compared to Echo Show, and the merits of the small form factors of Echo Dot were outlined by half of participants. 
For example, P2 emphasized: 
{\it ``the Dot is smaller and more inconspicuous. So it's easier to fit into a smaller space. I like the convenience of that [...] The Dot is less intrusive in your apartment so that you can put it in different places more conveniently [...] I like the size of the Dot and the discreet shape of it''}.
After setting up the Echo Show, P12 commented: {\it ``[the Echo Show] would be too bulky! For me, you could see, I don't have much space on the table. I have a small apartment, and I don't have a lot of spaces or things. [...] For something that is new, you have to think about it, adjust it, and learn it''}.
Despite this, few participants suggested while the bulky size might affect the experience for setting up the Echo Show, it will not affect the long-term uses.
For example, {\it ``the Echo Show is bulky, but I would just leave it there and don't move it around''}~(P15).

\subsection{RQ2: How Does the Built-In Touchscreen Affect the Older Adults' Behaviors and Experience of Conducting Self-Report Daily Diary Survey?}\label{sec::results::ema}

\begin{figure}
    \centering
    \includegraphics[width=0.47\textwidth]{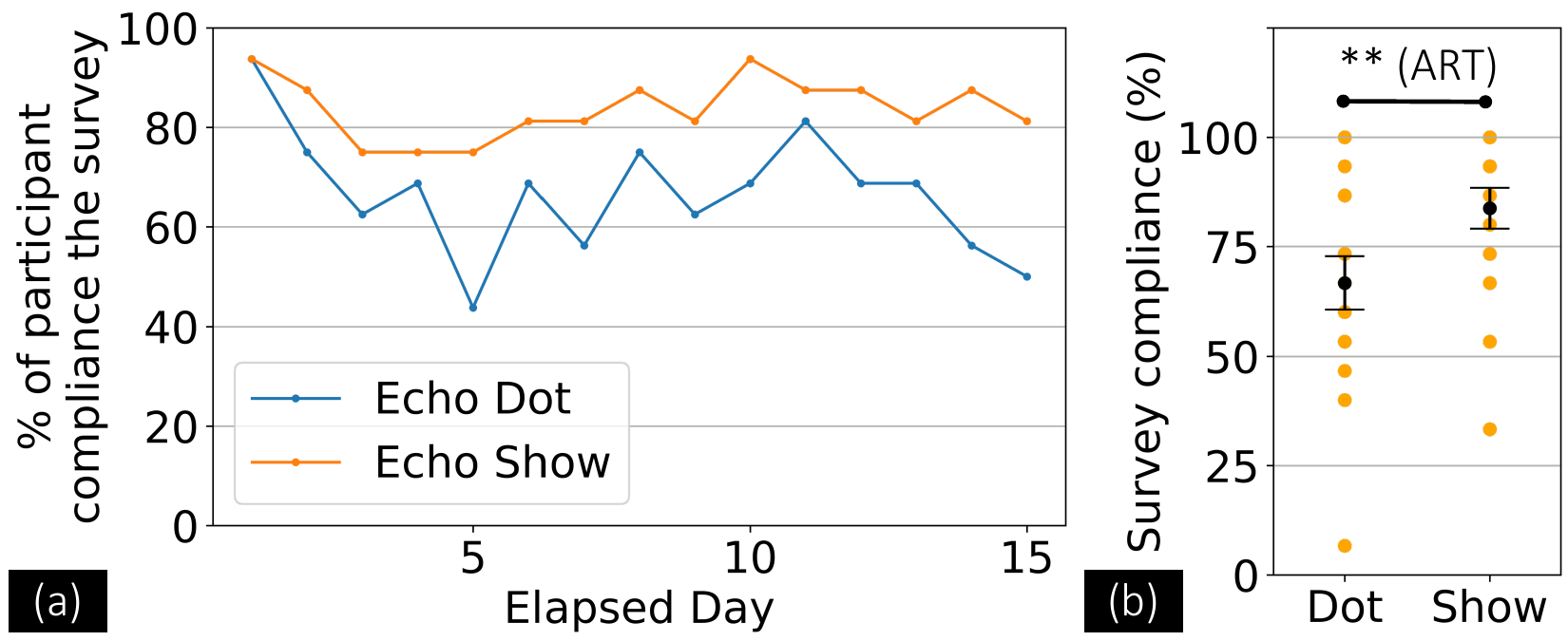}
    \vspace{-0.1in}
    \caption{Evaluations of survey compliance. (a) \% of participants that complied with the survey on each day; (b) Survey compliance rate on Echo Dot and Echo Show; Notations for indicating the statistical significance of \posthoc~test: * = $ .05 > p \ge .01$, ** = $ .01 > p \ge .001$, *** = $p < .001$.}
    \Description{Figure 7a shows two line plot shows that demonstrate the percentage of participant compliance the survey at each experimental day. The Echo Show exhibited slightly higher percentage of participant compliance the survey starting from the second day. Figure 7b describes the average survey compliance measured after using the Echo Dot and the Echo Show. Statistical significance is detected, and is annotated as ``*** (ART)''.}
    \vspace{-0.2in}
    \label{fig::ema_compliance}
\end{figure}

\begin{figure*}
    \centering
    \includegraphics[width=\textwidth]{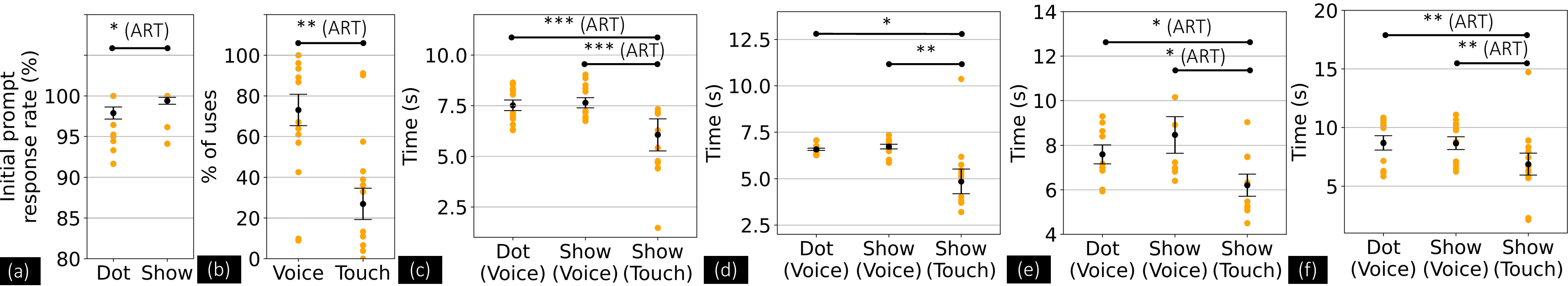}
    \vspace{-0.25in}
    \caption{Evaluation results: (a) Initial prompt response rate for the \emph{number}-type question, (b) \% of uses of voice \vs~touch while using Echo Show to complete the daily survey; (c) Overall response latency; (d - f) Response latency of \emph{binary}- (d), \emph{number}- (e) and \emph{``Likert''}- (f) type questions. Notations for indicating the statistical significance of \posthoc~test: * = $ .05 > p \ge .01$, ** = $ .01 > p \ge .001$, *** = $p < .001$.}
    \Description{Figure 8a shows the initial prompt response rate for number type question measured after using the Echo Dot and the Echo Show. Statistical significance is detected, and is annotated as ``* (ART)''; Figure 8b shows the percentage uses of voice and touch input modality while using the Echo Show. Statistical significance is detected, and is annotated as ``** (ART)''. Figure 8c describes average response latency for EDV, ESV, and EST. The statistical significance between EDV and EST is annotated as ``*** (ART)''. The statistical significance between ESV and EST is annotated as ``*** (ART)''. Figure 8d demonstrate average response latency for EDV, ESV, and EST. The statistical significance between EDV and EST is annotated as ``*''. The statistical significance between ESV and EST is annotated as ``**''. Figure 8e shows average response latency for EDV, ESV, and EST. The statistical significance between EDV and EST is annotated as ``* (ART)'', and the statistical significance between ESV and EST is annotated as ``* (ART)''. Figure 8f describes average response latency for EDV, ESV, and EST. The statistical significance between EDV and EST is annotated as ``** (ART)'', and the statistical significance between ESV and EST is annotated as ``** (ART)'';}
    \vspace{-0.15in}
    \label{fig::ema_quant}
\end{figure*}

\begin{figure*}
    \centering
    \includegraphics[width=\textwidth]{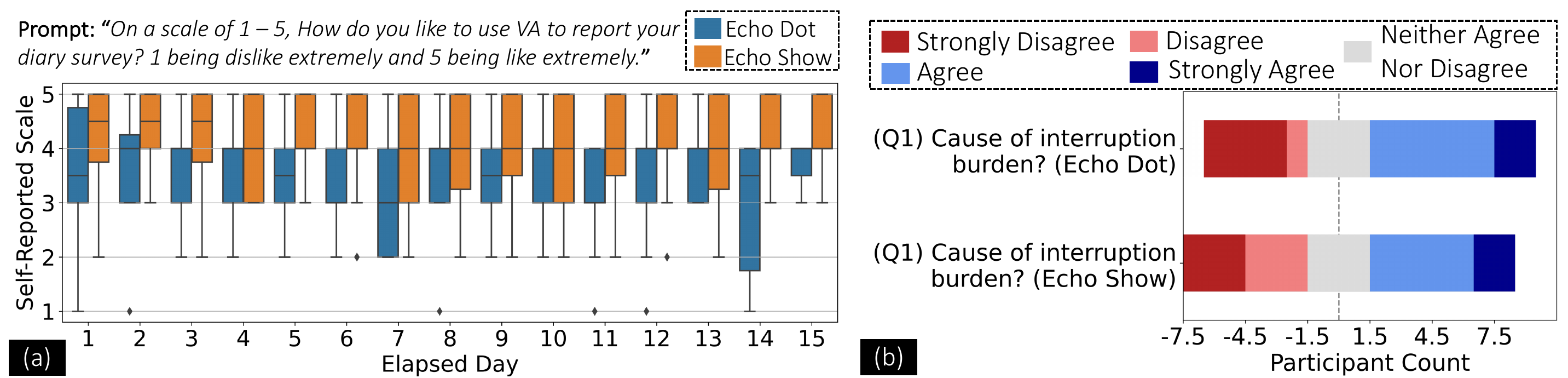}
    \vspace{-0.25in}
    \caption{Evaluation results of (a) participants' responses of the usability prompt \emph{``on a scale of 1 – 5, how do you like to use the voice assistant to report your diary survey? 1 being dislike extremely and 5 being like extremely''} reported through VAs in each study day, and (b)~the $5$-point Likert survey of how participant think the journaling of diary could cause interruption burden to the daily life routine (Q1). Incompleted responses were excluded from (a).}
    \Description{Figure 9a shows the box plot of participants’ responses of the usability prompt ``on a scale of 1 - 5, how do you like to use voice assistant to report your diary survey? 1 being dislike extremely and 5 being like extremely'', reported through VAs in each experimental day. Incomplete responses were excluded. A slightly higher ratings of the Echo Show are exhibited. Figure 9b describes the 5-point Likert survey results of how participant think the journaling of diary could cause interruption burden to the daily life routine. Slightly more participants believed that using the Echo Show will not cause interruption burden to their daily life.}
    \vspace{-0.15in}    \label{fig::ema_disturbance_overall}
\end{figure*}

Overall, participants demonstrated the usefulness of conducting such voice-first daily diaries (\eg~{\it ``it kind of reminded me to eat more fruit and many other aspects to keep myself healthy. So I thought it was helpful, just like a memory enforcement, as I don't think it sometimes''}~(P14)).
First, we demonstrate a statistical significance of the survey compliance rate (ART: $F_{1,15} = 10.33$, $p = .006$, $\eta_p^2 = .41$, Fig.~\ref{fig::ema_compliance}b) while using the Echo Dot and Show.
Fig.~\ref{fig::ema_compliance}a further demonstrates the \% of participants who complied with the diary over each of $15$-day session, where participants using Echo Show exhibited a slight higher compliance rate from the second day, compared to the voice-only alternative.
Second, while no statistical significance was detected in terms of question completion rate (ART: $p = .600$) and initial prompt response rate (ART: $p = .610$), a weak statistical increase of initial prompt response rate was captured (ART: $F_{1,15} = 5.12$, $p = .039$, $\eta_p^2 = .25$, Fig.~\ref{fig::ema_quant}a).
Third, while using Echo Show for journaling diaries, participants adopted voice as the input modality, more frequent than that of touch input (ART: $F_{1,15} = 9.91$, $p = 0.007$, $\eta_p^2 = 0.40$, Fig.~\ref{fig::ema_quant}b), with an average of $73.06\%$ versus $26.94\%$. 
Despite this, among $12$ participants (except P3, P10, P11, and P15) who have used touch input, we found that using touch input enabled a shorter response latency significantly, compared to the voice counterpart (through Echo Dot or Show) (ART, $F_{2, 34} = 14.40$, $p < .001$, $\eta_p^2 = 0.56$, Fig.~\ref{fig::ema_quant}c), with the average responses of using touch input being $6.056$ seconds versus $7.639$ seconds and $7.513$ seconds using voice input by Echo Show and Echo Dot respectively.
A similar observations have been captured in terms of measured response latency for {\it binary}- ($F_{2,33} = 7.08$, $p = .003$, $\eta_p^2 = .30$, Fig.~\ref{fig::ema_quant}d), {\it number}- (ART: $F_{2,16} = 5.16$, $p = .019$, $\eta_p^2 = .39$, Fig.~\ref{fig::ema_quant}e), and {\it Likert}- (ART: $F_{2,22} = 9.63$, $p < .001, \eta_p^2 = 0.47$, Fig.~\ref{fig::ema_quant}f) type questions.
Fourth, we found a higher median of preference rating of using Echo Show to report diary survey compared to the voice-only counterpart by analyzing self-reported Likert scale of the usability prompt shown in Fig.~\ref{fig::ema_disturbance_overall}a.
Finally, Fig.~\ref{fig::ema_disturbance_overall}b demonstrates that slightly more participants believed that using Echo Show will {\it not} cause interruption burden to their daily life, compared to using Echo Dot.
Through qualitative analysis, we identified four findings.

\vspace{4px}
\noindent \textbf{Touch input is faster, but responding via voice is still preferred.}
Most participants appreciated the merits of hands-free interactions using voice to journal diaries~(\eg~{\it ``it was interesting trying to [keep health diaries] when you weren't sitting right at the device [the desktop PC] and it turned out that I had to do it using paper and pencil. Whichever devices or systems [VAs] that will help you do that would be very valuable''}~(P8)).
However, nearly all participants subjectively believed that inputting response by touch could be faster compared to using speech, which verified the validity of Fig.~\ref{fig::ema_quant}c. 
Some participants chose to use touch to interrupt the delivered prompts.
Testimonies include: {\it``with the Show, I don't need to listen to the whole description. As soon as it is displayed, I know what to answer. I can move through the script faster''}~(P4), {\it ``having a touchscreen is faster. If I see something; I touch it [to submit my responses]; And it will go to the next question. If everything has to be oral, it has to ask me before I could answer. So with the touchscreen, it is a lot faster for things like going through checklists''}~(P13), and {\it ``response by touching speeds up quite a bit!''}~(P14).

However, participants overall responded more often to the prompts via speech than touch~($73.06\%$ \vs~$26.94\%$), and this was confirmed by testimonies such as {\it``I didn't use the touchscreen for any general use, I just use the voice for interactions.''}~(P4).
\textbf{Out of arms' reach} and \textbf{inconveniences caused by impaired mobility} are the most common reasons discussed among participants.
For example, {\it``speaking is easier because you don't have to lean over to press the buttons''}~(P1, with significant hand tremors) 
{\it ``if I am standing and I can reach [to the touchscreen]. I can respond quicker with a touch than I can with saying something and waiting for [the Echo Dot] to come back to me [through speech] with a question [...] that's a lot faster for me to read than for me to listen to Alexa talking about it''}~(P10),
and {\it ``usually I would be 10 feet away instead of  have to be right next to it''}~(P12). 
P13 also appraised the voice over touch based on his past observations: 

\begin{quote}
    {\it ``If you can do it without walking over to it. That's great! But if you have to walk over to it [to make a touch response], it can be a great hardship. [...] I was in the care center yesterday and there's a guy who's a 93 years old man who had surgery on his shoulder and his hip. He fell down and broke both his shoulder and his hip. He can't get up and touch the screen. So for him, something that he could operate with just voice would be very important. So I think probably around 85 years old is when that starts to become an issue, the issue of get up and go, and touch the display, rather than interact verbally!''}~(P13)
\end{quote}

This insights echos P9's comment, who needs a wheeled walker for moving: {\it ``you have to be up close to the screen to really touch the button. Whereas, in terms of the Dot, you could be 15 or 20 feet away, and that's not an issue''}.
Few participants also mentioned the \textbf{unpleasant visual experience due to discernible splotches caused by finger touch}
For example, {\it``when you have a touchscreen and touch it all the time it gets splotchy. So if there's an alternative, like using your voice, sometimes I just simply prefer to do that.''}~(P2).

\vspace{4px}
\noindent \textbf{Responding by speech needs more support for controlling the conversation flow.}
While using speech as the input modality, all participants emphasized that the flow of conversation should be {\it``more interactive and conversational''} (P1).
%
% First aspects
Participants identified the need for accepting longer responses (\eg~{\it ``I wanted to expand in an answer but there was no way of doing that''}~(P7)) and the short response time limit was compounded by the tendency of some participants to repeat the question at the beginning of their response. For example, without touchscreen, P8 sometimes exhibited the following interaction pattern, causing the system to fail in capturing valid diary responses:

\begin{quote}
    {\bf Echo Dot}: {\it ``How many hours did you walk outside today?''} \\
    {\bf P8}: {\it ``How many hours ... [unconsciously repeating the prompt causing the failures of valid response capturing]''}
\end{quote}

Additionally, due to the ambiguous nature of the speech conversation, participants suggested the need of designing additional ways to control the flow of the questionnaires ({\it ``I wanted something more interactive that we could go back and forth with kind of a conversation.''} (P1)). 
Participants particularly wanted to have a way to go back and revise previous responses (\eg~{\it``[it should] allow me to correct answers I've already said before''} (P8)).

\vspace{4px}
\noindent \textbf{Visual output could be helpful for information consumption.}
While the voice has been adopted as the major way for inputting information among older adults participants, many participants suggested the usefulness of having visual elements for information consumption, which might be one possible reason for Fig.~\ref{fig::ema_disturbance_overall}a.
One reason of such usefulness is that the question texts that are persistently shown on the touchscreen could help with older adults' short working memory, despite that the designed diary questions could be easily responded fast.
For example, {\it ``being old means my memory is shorter. By having the visual, it's easier to keep moving along [while reporting the daily diaries]. Whereas with the Dot, you think you're gonna say things, but then you forget the questions [...] With a visual. You have it [the diary questions] there and it keeps your mind focused on what you want to do or say''}~(P15).
Similarly, P12 also mentioned the usefulness of seeing possible Likert responses on the display: {\it ``I liked [the touchscreen], when the device asked a question and then it showed `one', `two', `three', `four' and `five', and you could see. It's easier for me to know and to remember. Whereas with the Dot, you have to remember what [the device] said, and sometimes [after the prompt being announced] I'll think, oh.. what range did the Alexa tell me''}.

\vspace{4px}
\noindent \textbf{While integrated reminders are needed for both devices, the larger form-factors of the Echo Show could lead to higher diary compliance.}
We encouraged participants to choose any methods they preferred for reminding themselves to conduct the daily diary survey.
$13$~participants reported that they remembered {\it ``just based on memory''} or {\it ``memorized the task as part of their daily routines''}, two participants used the reminder features on the standalone VAs or their smartwatch, one participant simply used his notes and calendar.
While all participants initially felt confident about their selected reminders before the study, it turned out that participants still forgot. For example: {\it`` I thought it would be easy for me to remember to do it around meal times every evening because that's when we get together in the kitchen. But unfortunately, I didn't always remember to do that''}~(P10).
Implicitly, P13 explained a possible reason of forgetting journaling daily diaries while the Echo Dot being covered by papers:
{\it ``I just put my Alexa on the desk and I usually [completed the daily diaries] as long as I saw it. But there were couple of days missing when my desk was super messy and had my Echo Dot covered up by papers''}~(P13).
This could be one possible reason of the decreasing of compliance rate while using Echo Dot (Fig.~\ref{fig::ema_compliance}), where forgetting is one major reason that cause failures of survey compliance.

\begin{figure*}[t]
    \centering
    \includegraphics[width=\textwidth]{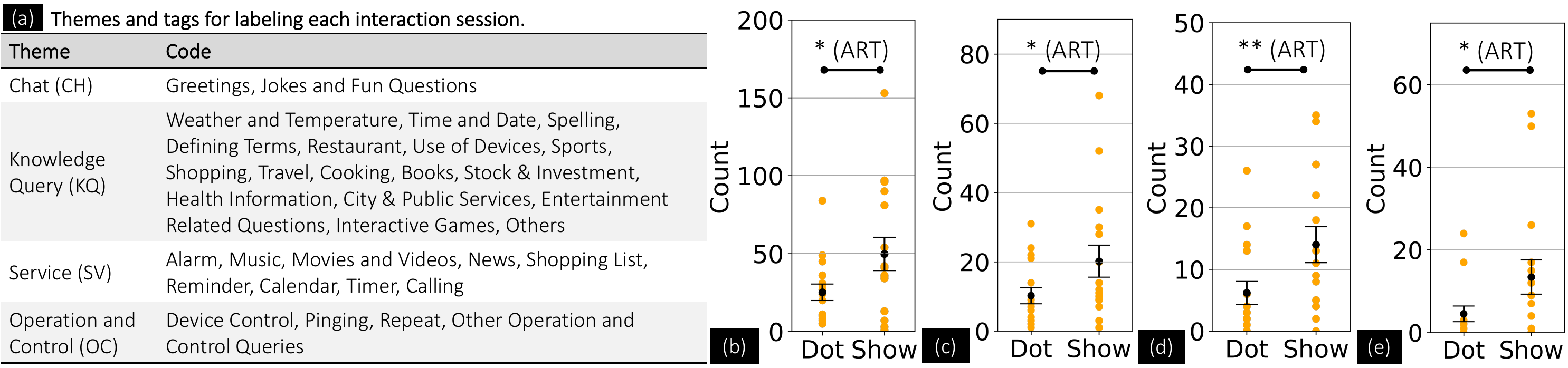}
    \vspace{-0.25in}
    \caption{Characterizations of general uses. (a)~Themes and codes used to label each interaction session. (b)~The total number of interaction sessions being measured for each participant during the study for both devices. (c - e) The total number of interaction sessions related to Knowledge Query (KQ) (c), Service (SV) (d), and Operation and Control (OC) (e) while using both devices. Notations for indicating the statistical significance of \posthoc~test: * = $ .05 > p \ge .01$, ** = $ .01 > p \ge .001$, *** = $p < .001$.}
    \Description{Figure 10a lists the themes and tags used to label each interaction session. First row includes the theme of chat (CH), with the generated tags including: greetings, jokes and fun questions; Second row includes the theme of knowledge query (KQ), with the generated tags including: weather and temperature, time and date, spelling, defining terms, restaurant, use of devices, sports, shopping, travel, cooking, books, stock & investment, health information, city & public services, entertainment related questions ,interactive games, others; Third row includes the theme of service (SV), with the generated tags including: alarm, music, movies and videos, news, shopping list, reminder, calendar, timer, calling. The final row includes the theme of operation and control (OC), with the generated tags including: device control, pinging, repeat, other operation and control queries. Figure 10b shows the total number of interaction sessions being measured for each participant over the study session for the Echo Dot and Echo Show. The statistical significance is detected with ``* (ART)'' being annotated. Figure 10c demonstrates the total number of interaction sessions related to knowledge query. The statistical significance is detected with ``* (ART)'' being annotated. Figure 10d describes the total number of interaction sessions related to service. The statistical significance is detected with ``** (ART)'' being annotated. Figure 10e shows the total number of interaction sessions related to operation and control. The statistical significance is detected with ``* (ART)'' being annotated.}
    \vspace{-0.10in}
    \label{fig::general_device_factor}
\end{figure*}

\begin{figure*}[t]
    \centering
    \includegraphics[width=0.80\textwidth]{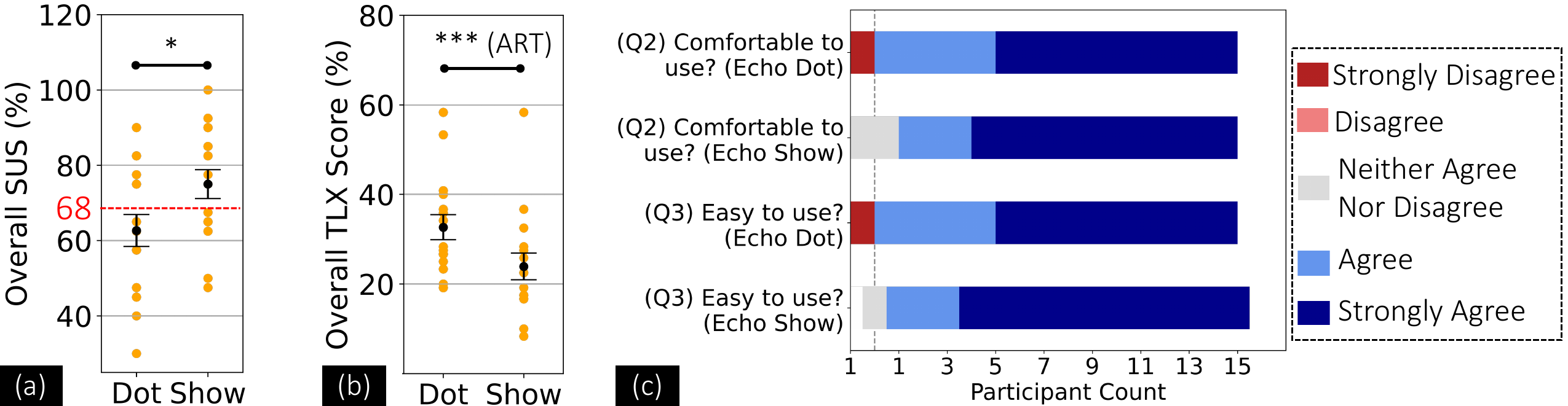}
    \vspace{-0.1in}
    \caption{(a) Overall SUS~\cite{Brooke1996, Brooke2013} and (b) overall NASA TLX~\cite{nasatlx} scores. A higher SUS and lower TLX score imply a better user experience. Notations for indicating the statistical significance of \posthoc~test: * = $ .05 > p \ge .01$, ** = $ .01 > p \ge .001$, *** = $p < .001$. (c) The 5-point Likert survey results of how strong participants agree with that the device is (Q2) \emph{comfortable to use} and (Q3) \emph{easy to use}. }
    \Description{Figure 11a describes the overall SUS of the Echo Dot and Show. The statistical significance is detected with ``*'' being annotated. Figure 11b demonstrates the overall NASA TLX of the Echo Dot and Show. The statistical significance is detected with ``*** (ART)'' being annotated. Figure 11c describes the 5-point Likert survey results of how strong participants agree with that the device is (Q1) comfortable to use, (Q2) easy to use, and (Q3) could cause interruption burden to daily routine.}
    \vspace{-0.15in}
    \label{fig::questionaire_results}
\end{figure*}

\subsection{RQ3: How Does the Built-In Touchscreen Affect the Older Adults’ Behaviors and Experience of using VAs for General Purposes?}\label{sec::results::gu}
\noindent 
Our coding approach generated $31$ codes categorized into four themes (Fig.~\ref{fig::general_device_factor}a).
Notably, while using the {\it Calling} service is out of our scope due to the needs of importing participants' contact books into our experimental accounts, which is not allowed by our IRB due to privacy concerns, we captured the intention of using such service by P13 and P16, leading to failure responses.
Such findings were also verified by P13's questions during interviews of Phase 3: {\it ``can I make a phone call using Alexa? Can I say Alexa, call my sister Rose? [...] I can do that after the study, right?''}~
Overall, we demonstrate a weak statistically increasing of measured interaction session while using the Echo Show compared to the Dot (ART: $F_{1,15} = 5.01$, $p = 0.041$, $\eta_p^2 = 0.25$, Fig.~\ref{fig::general_device_factor}b).
While no statistical significance were observed for interaction sessions related to {\it Chat}, increases of {\it Knowledge Query} (ART: $F_{1,15} = 6.88$, $p = 0.019$, $\eta_p^2 = 0.31$, Fig.~\ref{fig::general_device_factor}c), {\it Service} (ART: $F_{1,15} = 9.82$, $p = 0.007$, $\eta_p^2 = 0.40$, Fig.~\ref{fig::general_device_factor}d), and {\it Operation and Control} (ART: $F_{1,15} = 6.14$, $p = 0.026$, $\eta_p^2 = 0.29$, Fig.~\ref{fig::general_device_factor}e) were observed.
As for participants' self-reported survey results, we demonstrate a statistically higher overall SUS score~($F_{1,30} = 4.67$, $p = 0.040$, $\eta_p^2 = 0.13$, Fig~\ref{fig::questionaire_results}a) and a lower overall TLX score~(ART: $F_{1,15} = 29.20$, $p < .001$, $\eta_p^2 = 0.66$, Fig.~\ref{fig::questionaire_results}b) while using the Echo Show compared to Echo Dot, implying a better overall user experience.
In particular, we measured the average overall SUS score yielded by using Echo Show being $75\%$, which is empirically considered as a good rating~\cite{Sauro2011}.
Fig.~\ref{fig::questionaire_results}c shows that while most participants were comfortable and felt it was easy to use Echo Dot and Show, one participant held negative opinions.
The qualitative analysis outlines our findings from three perspectives.

\vspace{4px}
\noindent \textbf{Participants preferred the lower disturbance level afforded by Echo Dot compared to Show.}
Participants reported how touchscreen caused visual disturbances during times of non-use: {\it ``[...] the screen changes all the time, and that can be irritating.''}~(P2).
While time displayed on screen was recognized to be useful, older adults emphasized that most visualizations could cause disturbances to some extent (\eg~{\it``I would prefer to only show the time until I actually asked a question. But Amazon has prevented that from happening because they want to show you ads and other kind of things''} (P2)).
Further along this theme, two participants mentioned that the brightness of the display might affect the sleep quality when placed in the bedroom:
~{\it``the Echo show is pretty bright. If you have it in your bedroom, I had to turn it toward the part of my desk [to avoid direct light]''} (P6), {\it``[the Echo Show] is on my nightstand, and I couldn't figure out how to control [...], so I got to turn that around to face another way''} (P1).

\vspace{4px}
\noindent \textbf{Participants enjoyed seeing the additional auxiliary visual component on the display.}
Participants enjoyed the visual output together with the audio responses when using the device for general purposes.
Examples include {\bf auxiliary visual elements that are not announced by the voice output} (\eg~{\it ``I like seeing the responses to questions. If you ask the Show, to add two numbers or multiply two numbers, it actually displays as well as telling you the answer [...] If you ask it about the weather. It'll tell you what the weather is going to be but it'll also show you little symbols [...]. So you get additional pieces of information from the screen that you don't get from the device without the screen''} ~(P2)), {\bf heterogeneous feature suggestions} (\eg~{\it ``the Echo Show is more preferable, because it's making suggestions about how it can be used''}~(P15)), {\bf persistent visual information such as time and weather} ({\it ``the Echo Dot just sat there. Whereas the screen of the Echo Show gave me information about the day, the time, the weather, and also hints! It was like having a companion in the house that was silent, but still there! [...] I found it much more helpful and much more enjoyable as more than a simple device! [...] I enjoy getting up each morning and having it refresh me on the date, the time, and the weather, whereas for the Dot, I would have to ask them to give me the information. I guess I'm sort of lazy and prefer having it all out there for me''}~(P16)), and {\bf explicit visual outputs for greetings and creating a sense of companionship} ({\it ``when I said `thank you' to Alexa, there was a little blurb on the screen with the thank you message. That made me laugh and made me have a sense of companion''}~(P9)).

Many participants explained how visual outputs could make their generic uses of VAs become easier.
First, most participants emphasized the usefulness of having visual components for specific types of information consumption.
For example: {\it ``I love the display while listening to music because I can see the lyrics''}~(P5), {\it ``putting something [P9 added that the most medias experienced were Youtube videos] on screen to entertain me while cooking''}~(P9) and {\it ``when I wanted to see a recipe, the Echo Dot could not do that! It's a written thing that's laid out for me. But for Echo Show, it gives us a screen which can show a list of items. And that's one very useful way we could understand a recipe'' }~(P10).
Second, some participants believed the visual elements could help on reinforcing their memory regarding the output voice information. 
For example, {\it ``I didn't pay much attention to the screen during most of time, but I think it is always useful to have visual output to reinforce of what you're hearing''}~(P14).
Finally, additional visual output might also compensate the hearing impairment though the visual sensory experience.
For example: 
{\it``if I am not close to an Alexa device, I don't always hear the words that she says. [With Show], the words will appear on the screen as well. So I can look at it, as well as hear it. [...] For people who have hearing aids and who don't necessarily wear them at home, Alexa can be very difficult to understand if you're not standing right next to her''}~(P2).

\vspace{4px}
\noindent \textbf{Impacts of failures of speech recognition.}
While participants enjoyed the conversational capabilities brought by voice $+$ visual output, our results identified the impacts of ambiguity and recognition failures of voice commands. 
While using Echo Dot, P3 complaint: 
{\it ``occasionally, when I said `Alexa! stop music', it would not stop the music. I had to unplug it and it was frustrated''}~(P3).
Despite this, we found many participants subjectively think such occasional speech recognition failures will not cause significant impacts on the overall user experience.
For example, {\it ``it happened about maybe 10\% to 20\% of the time [refer to the time when VAs cannot understand P13's intent] [...] It's usually because I talk too fast or my talking are not clear or something like that. It's not the devices fault [...] I think errors are mind and not about the machines''}~(P13).

\vspace{4px}
\noindent \textbf{The location and placement of the display affect what and how to use it.}
While some participants conceptually acknowledged the usefulness of visual output and touch input, they also suggested how incorrect placement of the device could reduce its usefulness.
During one focus group, P8 reasoned:
{\it``my device was stuck in a corner that was close to an outlet, and that was not particularly convenient [to access]''}. 
P1 emphasized that being able to have hands-free interaction with the device was much more useful compared to the touch modality, possibly due to the non-optimal device placement, which could be one possible reason causing high physical and temporal task load in the post-study survey in Fig.~\ref{fig::questionaire_results}.
Example testimonies include: {\it``it depends on where [the Show] is located. It was not easy to do the touch and it was much easier to use voice, because I can do it a little bit from a distance.''}~(P1),
and {\it ``based on my experience so far, I would do it without touching. I'm lazy and it's just easier! We had the device being about six feet away! You saw my office. It's at the end of my desk and I sit over here in the corner, so it's a lot easier to just yell over, such as `what time is it' or `Alexa? Give me a 10 minute warning' or something like that. So I would probably use it more if it was like right in front of me but my desk is so full now''}~(P13).

\begin{figure*}[t]
    \centering
    \includegraphics[width=\textwidth]{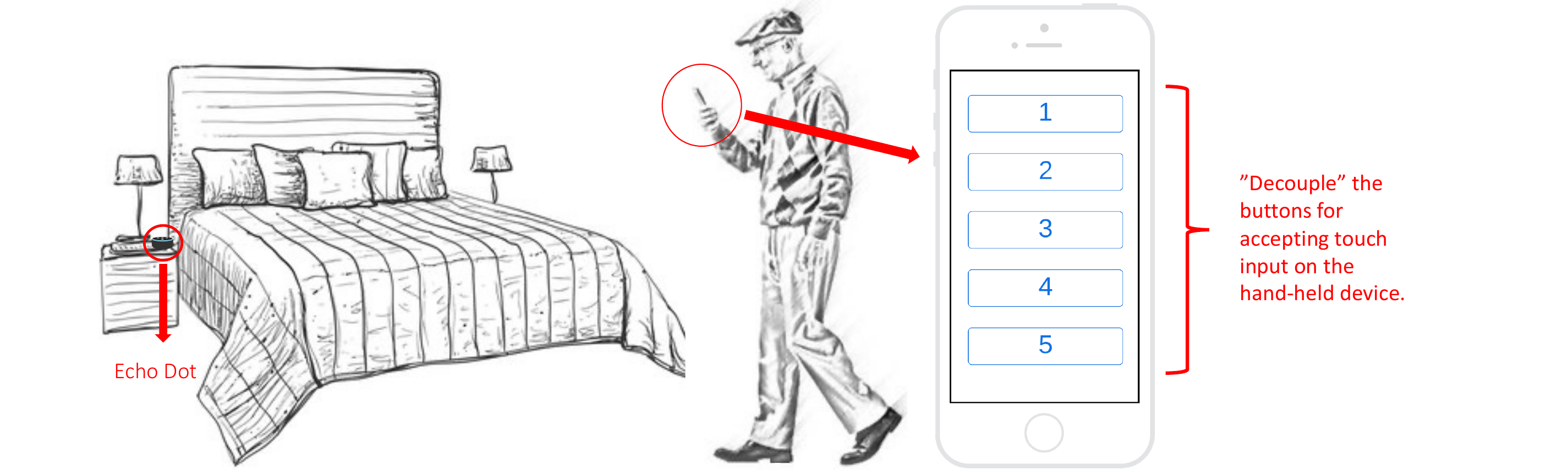}
    \vspace{-0.25in}
    \caption{Conceptual design to use touchscreen for better control. Instead of using a built-in touchscreens, a multi-device system could combine a voice-only VA with a hand-held device used as an additional channel for inputting responses.}
    \Description{Figure 12 shows the conceptual design sketch for achieving alternative designs to better support one-way modalities. Instead of using built-in touchscreens, a multi–device system could combine a voice–only VA with a hand–held device used as an additional channel for inputting responses. The Echo Dot is placed on the countertop next to the bed. And the older adult is holding a smartphone and ready to input responses through five buttons on the phone.}
    \vspace{-0.15in}
    \label{fig::future_modality}
\end{figure*}

\section{Discussion}\label{sec::discussion}

\subsection{Using the Touchscreen for Device Setup and General Uses}
Through our deployment study, we found that older adults held overall positive opinions for using the touchscreen as a secondary modality on top of voice. 
With such insights, we offer three key design opportunities.

\vspace{4px}
\noindent{\bf Integrating suggestions of device placement into the interactive device setup phases.}
The nature of \mbox{hands-free} and eyes-free interactions has pushed voice user interfaces as promising candidates for ambient assistive living \mbox{technology}, which eventually helps older adults better access and interact with inherently complex supporting \mbox{technologies}~\cite{Goetze2010, Kleinberger2007}. 
While introducing a built-in touchscreen might enhance robustness and usefulness of the interaction system, the device accessibility might be degraded, as the interactions with touchscreen are not fully hands-free and eyes-free.
Some older adults (\eg~P1, P8, and P11) pointed out that the placement of the devices could affect the general uses.
Although being instructed to place the device in a preferred location, nearly all participants only considered the outlets' location and the size of the device (\eg~{\it``the cable is too short! I could only put it right here [points to an awkward place that is hard to reach]''}~(P7)). 
%
% problem
While setting up today's voice $+$ touchscreen VAs, there is no information regarding the potential degradation of the touchscreen-related experience caused by non-optimal placement of devices.
Our finding implies that relying on only users' intuition to place the voice $+$ touchscreen VAs might not be feasible and could diminish the values of the touchscreen.
Therefore, one future improvement could be to design ways to help users decide on device placements during the setup phase.
Common features and suggested placement locations could also be crowd-sourced and provided to help older adults make the best decision. 

\vspace{4px}
\noindent{\bf Maximizing the merits of visual components.}
Provided finding a suitable power outlets, we demonstrated a strong preference of participants to place their devices in an optimal location to maximize the accessibility of visual components~(\eg~P2 put the Echo Show in front of her workstation, Fig.~\ref{fig::teaser}h, j). 
We also discussed how older adults appreciated the companion visualizations of the icons and texts of audio responses, during both diary journaling and generic uses. 
While Hu~\etal~\cite{Hu2022} suggested that some older adults intended to bypass voice output and treat such voice-first VAs as touchscreen-only device (\eg~tablet) when it comes to incorrect speech recognition~\cite{Whitenton2017}, we showed that older adults could, and are willing to, consume the {``voice-first output''} for general uses.
While Han~\etal~\cite{Han2022} showed the promising of using touchscreen to visualize EMA data through a preliminary interview study, it is still unclear how to instantantiate such design tenet.
Along this findings, future design might measure and consider {\it how}, {\it what}, and {\it when} to visualize, in order to maximize efficiency while consuming such {``voice-first output''}.

\vspace{4px}
\noindent{\bf Design opportunities for context-awareness.}
While we showed how the touchscreen increases efficiency in terms of information consumption, some older adults emphasized the setbacks of disturbance, caused by, for example, the brightness of the screen at night and displaying irrelevant content at the focus time.
Addressing such problems requires VAs that have the ability to adapt visualizations to real-world contexts.
Example designs include dimming (or turning off) the display when older adults are sleeping or not in the room, and keeping the home screen visualizations consistent during focusing hours (\eg~only show the virtual clock).
One future opportunity is to investigate how to detect and achieve context-awareness using less privacy-invasive sensors (\eg~light sensors, instead of cameras). This has also been previously suggested as one important concern by older adults who are adopting voice-first VAs~\cite{Bonilla2020, Chen2021}.

\subsection{Using the Touchscreen to Keep Health Diaries}
We showed that touchscreen affordances help older adults while journaling diaries, especially in terms of reducing response latency and increasing survey compliance.
These findings lead to three key design implications.

\vspace{4px}
\noindent {\bf Leveraging the affordances of touch input by decoupling primary and secondary input modalities. }
We showed that touch input leads to $\sim20$\% reduction in response latency, yet most participants still preferred to use speech input, mainly because the touch input was not always an accessible input modality.
This implies that the built-in touchscreen supports older adults only for information consuming, instead of information input.
Future design could focus on decoupling the input and output modalities. 
Fig.~\ref{fig::future_modality} illustrates an example where voice $+$ touchscreen VAs can be extended into a multi-device system that spans across user-attached (\eg~handled device) and user-detached devices (\eg~voice-only VAs).
While similar ideas have been proposed in TandemTrack~\cite{Luo2020} targeting general users who wanted to track exercise behavior, the differences between young and older adults in terms of technology proficiency and daily needs could pose unique challenges and needs to be investigated.
For example, creating such a system requires designers to consider the implicit increment of interaction complexities and the impacts of device maintenance tasks (\eg~charging and configuration), which might in turn lead to poor usability.
One future direction is to investigate the design trade-offs between enhanced accessibility of touch input and how usability could be sacrificed by the increased complexities brought by user-attached hand-held devices.

\begin{figure*}[t]
    \centering
    \includegraphics[width=\textwidth]{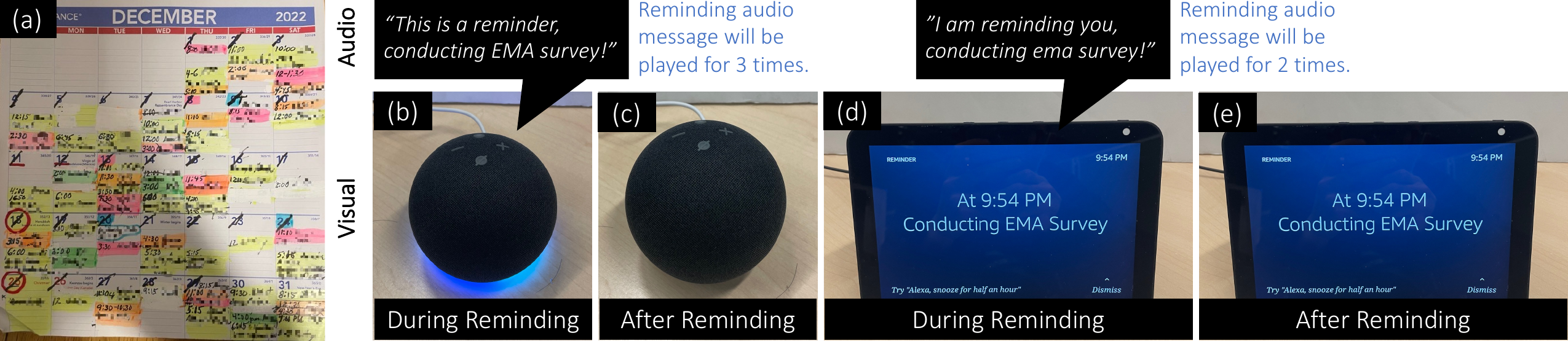}
    \vspace{-0.25in}
    \caption{(a) An example paper-based note for event reminding from P12. Reminder features on standalone Echo Dot (b - c) and Echo Show (d - e).}
    \Description{Figure 13a demonstrates the example calendar used by P12 for assisting reminders; Figure 13b - c shows example scenes that with Dot, the reminder message does not stay persistently after the initial triggering event; Figure 13d - e describes that with Echo Show, while the visual reminder message will stay, the proactive audio notifications are snoozed after the triggering event;}
    \vspace{-0.15in}
    \label{fig::reminder}
\end{figure*}

\vspace{4px}
\noindent{\bf Opportunities to use the touchscreen for better control. }
We showed that using voice input alone could cause ambiguity and incorrect speech recognition, in particular for diary responses, and using the touchscreen as a secondary input support is promising to address this issue.
%
% talk more succinct?
While Fig.~\ref{fig::future_modality} might address these challenges by supporting higher interactivity on a handled device, our observations during device setup indicate that most older adults found it confusing to set up VAs using a decoupled smartphone.
Therefore, instead of using voice-only VAs $+$ smartphone, a touchscreen might still be needed to ease such process, but does not necessarily need to be integrated in the VA. 
Fig.~\ref{fig::future_modality} illustrates a ``decoupled'' touch-device that can be used as an independent add-on to the VA.
Finally, introducing unnecessary features might increase a negative attitude among older adults~\cite{Mitzner2010}. 
To ease this problem, while still providing added functionality, future VAs should integrate more interactive guidance for helping older adults setting up devices and troubleshooting unexpected exceptions during device setup, and design metaphors for supporting more heterogeneous state navigation.

\vspace{4px}
\noindent{\bf Opportunities to use the touchscreen for diary reminders and beyond.}
Part of our study focused on diary studies that do not have strict time requirement for participants to conduct the survey.
Despite this, we asked participants to choose their preferred reminder methods for completing the survey.
Most participants were confident that they would simply remember the tasks just based on their memory, as it would become part of their daily routine.
However, we showed a different story: while measured survey compliance results ($\mu$, $66.67$\% and $83.75$\% for Dot and Show) are comparable to the existing $\mu$EMA systems where smartwatch vibration was used as the haptic notification~\cite{Intille2016} ($81.21$\%), our participants reported that they started to forget.
Designing effective reminding mechanisms for standalone devices is challenging due to the non-portability and the detached nature from end-users.
\mbox{Fig.~\ref{fig::reminder}b - c} show how without touchscreen, the reminder message does not stay persistently after the initial triggering event. Even on Show, where visual reminder messages will stay persistently on the display, the proactive audio message will be snoozed after the event \mbox{(see Fig.~\ref{fig::reminder}d - e)}. 
We, therefore, believe that leveraging a combination of standalone VAs with user-attached devices could be promising.
In an ecosystem like the one shown in Fig.~\ref{fig::future_modality}, proactive notifications could be designed on top of handheld devices without the need to set up a separate application.
Besides diary journaling reminders, future work might also investigate how to bring older adults calendars (\eg~Fig.~\ref{fig::reminder}a) into such multi-device ecosystem.

\section{Limitations}\label{sec::limitation}
We recognize three limitations that might hinder the applicability of our findings to a more generalized setting.
First,~although our study was performed in a naturalistic environment, we evaluated only $16$~older adults living in La Jolla in the United States.
This may lead to biases in experiences that different populations might not have engaged with.
Future work might investigate different groups of older adults who have a more varied experience with technology, live in different neighbourhoods, or share the VAs with their family members. 
Second,~although we only focus on standalone VAs, their uses introduced inherent setbacks when users are more mobile (\eg~travel frequently).
While assuming that older adults spend a considerable amount of time in their home, we did not assess interactions involving user-attached VAs when not at home. 
Future work could evaluate the ecosystem consisting of both user-attached and user-detached devices and their interactions. 
Third, our current study used Echo Dot and Echo Show as the testbed due to the dominant market share (Sec.~\ref{sec::methods::implementations}). 
While VAs from different vendors shared many similarities in terms of functions and designs, future work might consider evaluate other type of VAs in terms of older adults' perspectives.
Finally, part of our research focuses on general uses of standalone VAs. 
However, under the restrictions of our current IRB protocol, we were not allowed to link our experimental account with the third-party services that need participants' private data (\eg~ email and calling).
Future deployment could consider older adults' private accounts (under different IRB protocols), which might offer more insights on their behaviors.

\section{Conclusion}\label{sec::conclusions}
We conducted a within-subjects study (N = $16$) using the Echo Dot and Show to understand how the voice $+$ touchscreen VAs could influence older adults’ experience of device setup, diary journaling, and general uses.
Through a \mbox{$40$-day} \mbox{real-world} deployment, we found that during the device setup, older adults appreciated the advantages of the touchscreen, with the overall TCT reduced by roughly $50$\% when using Echo Show compared to Echo Dot.
As for diary journaling, while older adults enjoyed the visual output of touchscreen, they still preferred to respond to the prompts through speech, despite an approximately $20$\% of latency reduction while using touch input.
Finally, we found that touchscreens were effective in encouraging older adults to engage more with VAs for general uses, despite the fact that input through touch was still referred to as not senior-friendly by our participants.

\begin{acks}
    This work is part of project VOLI and was supported by NIH/NIA under grant R56AG067393. Co-author Michael Hogarth has an equity interest in LifeLink Inc. and also serves on the company's Scientific Advisory Board. The terms of this arrangement have been reviewed and approved by the UC San Diego in accordance with its conflict of interest policies. We appreciate insightful feedback from the anonymous reviewers and discussions with colleagues from The Design Lab at UC San Diego, including Matin Yarmand, Janet G. Johnson and Manas Bedmutha. We thank Christopher Han and Peng Wei Lee for the help on the early stage implementations, and Mary Draper along with residents from the Vi at La Jolla for the help on participant recruitment.
\end{acks}

\bibliographystyle{ACM-Reference-Format}
\bibliography{references}
\clearpage
\begin{appendices}
\section{Design of Diary Survey}\label{sec::app::ema_question}
\begin{figure*}[ht]
    \centering
    \includegraphics[width=\textwidth]{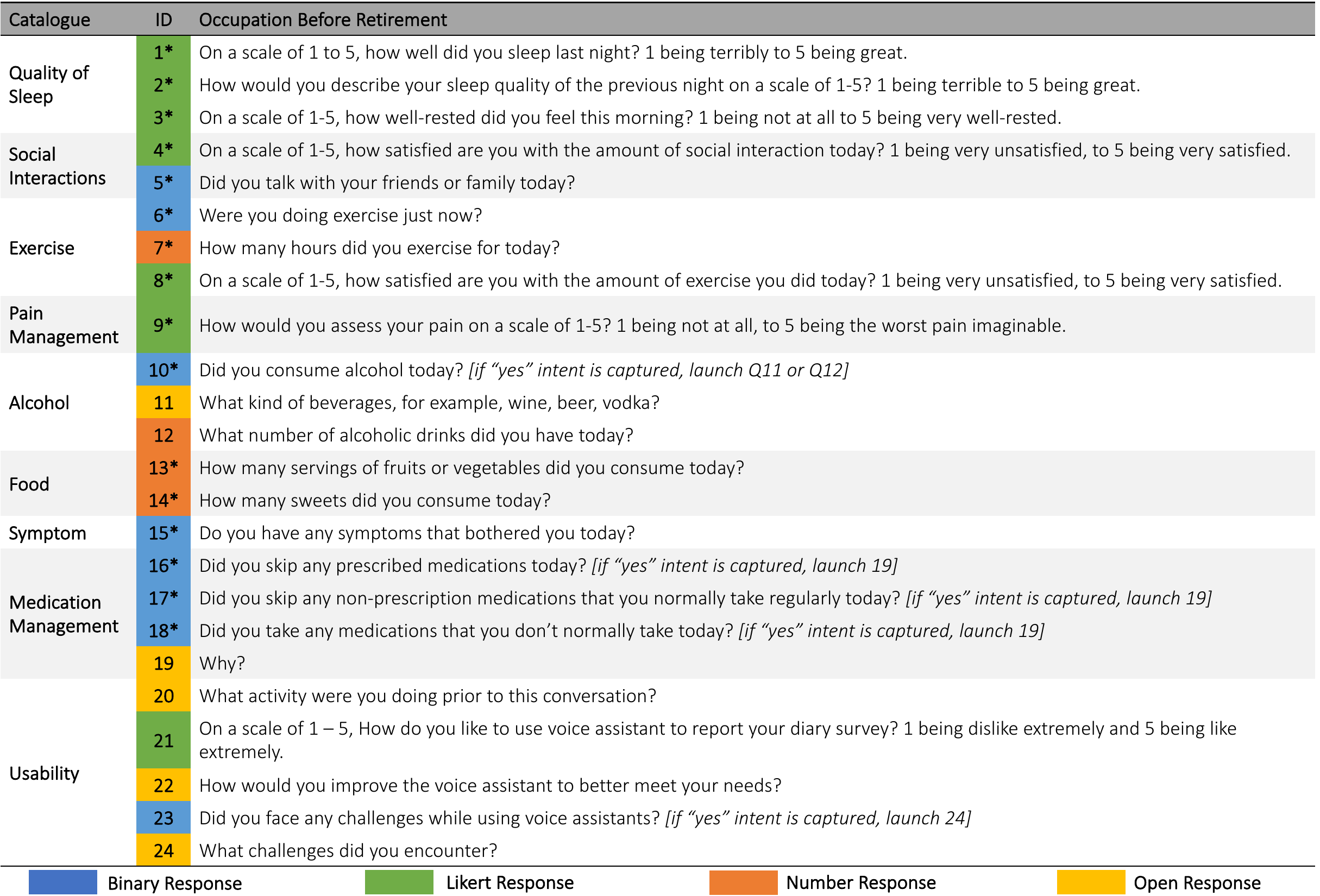}
    \vspace{-0.20in}
    \caption{We designed the diary survey focusing on eight themes of older adults' general wellness based on suggestions from professional geriatricians and empirical guidance from WHO~\cite{whoaging}. For each theme, we provided different paraphrased versions, and one question will be randomly chosen from those marked by ``*''. We also included the usability section to understand participants' \emph{in situ} user experience.}
    \Description{Figure 14: We designed the diary survey focusing on eight themes of older adults’ general wellness based on suggestions from professional geriatricians and empirical guidance from WHO. For each theme, one question will be randomly chosen from those marked by ``*''. We also include a usability section to understand participants’ in-situ user experience. Five types of expected answers are included in the study.}
    \label{fig:all_tasks_results}
\end{figure*}

This section provides supplementary details of the design of diary questions in Sec.~\ref{sec::method::ema}.
While establishing the validity of diary survey questions and the design of clinically-relevant diary studies is {\it beyond} our scope and left for future work, all older adults participants were instructed to provide their responses attentively as they will be carefully studied.
The goal was to ensure that the participants were actually spending efforts on carefully deciding the responses for each prompts, which eventually aimed to mock up a realistic real-world diary journaling experience.
We have explored and discussed iteratively the diary survey with one geriatric domain expert. Their focus during the design phase was to create prompts that could easily be answered by older adults, and could offer interesting insights into older adults' daily life for healthcare providers.
In addition to the eight themes centered around empirical guidance from the World Health Organization~(WHO)~\cite{whoaging}, we also added five questions at the end of each survey to understand older adults' \textit{in situ} user experience.  
Due to the anthropomorphic nature of conversational voice assistants, especially in terms of older adults users~\cite{Pradhan2019}, we emphasized a detailed disclaimer in the informed consent to not rely on the device for medical advice and we used {\it``If this is an emergency, call 9-1-1!''} as a short welcome message for each daily survey to remind older adults to not try to use the experimental testbed as a tool for seeking emergency help.
The full question set is displayed in Fig.~\ref{fig:all_tasks_results}.

\clearpage
\section{Guiding Questions for the Interviews}\label{sec::app::guiding_questions}
\begin{figure*}[h!]
    \centering
    \includegraphics[width=\textwidth]{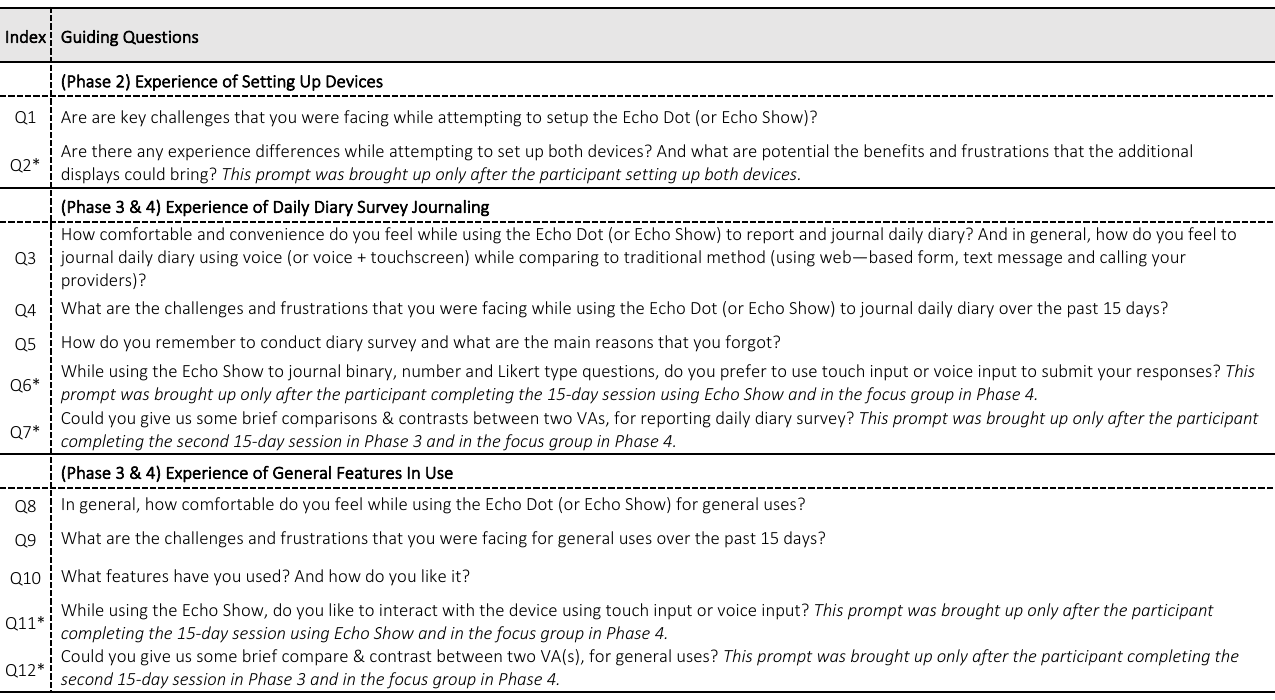}
    \vspace{-0.25in}
    \caption{Guiding questions used for semi-structured interviews at different study stages. Notably, the prompts Q2, Q6, Q7, Q11 and Q12, annotated by *, were only discussed at specific moments in the study stage.}
    \Description{Figure 15 shows the guiding questions used for semi-structured interviews at different study stages. 12 guiding questions are presented spanning across three themes.}
    \label{fig::app::guiding_questions}
\end{figure*}

This section provides supplementary details of the semi-structured interviews we conducted in Phase $2$ after device setup, Phase $3$ after each $15$-day session with Echo Dot and Show, as well as the focus groups in Phase $4$.
Fig.~\ref{fig::app::guiding_questions} shows the guiding questions we used for semi-structured interviews or focus groups at different study stages.
The responses and discussions to the guiding questions were expected to be open-ended and the participants were {\it not} expected to {\it only} answer the questions.
Instead, we encouraged participants to expand their responses and tell us more about their experience, stories, and rationales.

\clearpage
\section{Codebook and Themes from Interview Data Analysis}\label{sec::app::codebook}
\begin{figure*}[h!]
    \centering
    \includegraphics[width=\textwidth]{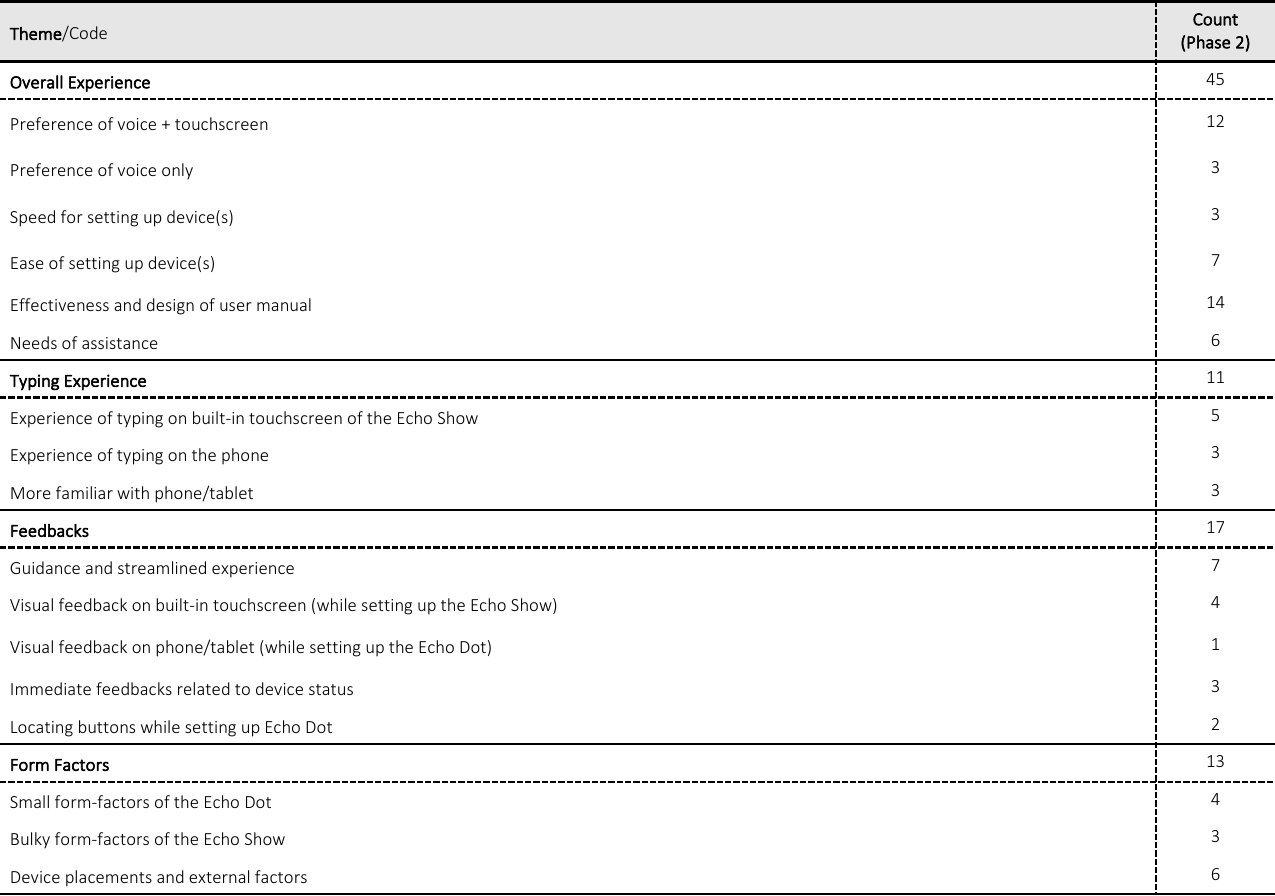}
    \vspace{-0.25in}
    \caption{The codebook resulted from our qualitative analysis of study Phase 2, showing four themes (bold). The ``Count'' refers to the number of participants' quote tagged with corresponding theme (or code). Notably, it is possible that more than one codes are assigned to a specific quote.}
    \Description{Figure 16: The codebook resulted from our qualitative analysis of study Phase 2, showing four themes (bold). The “Count” refers to the number of participants’ quotes tagged with corresponding theme (or code). Notably, it is possible that more than one code is assigned to a specific quote. Overall, 17 codes were generated across four themes: ``Overall Experience'', ``Typing Experience'', ``Feedbacks'', ``Form Factors''.}
    \label{fig::app::codebook_device_setups}
\end{figure*}

\begin{figure*}[h!]
    \centering
    \includegraphics[width=\textwidth]{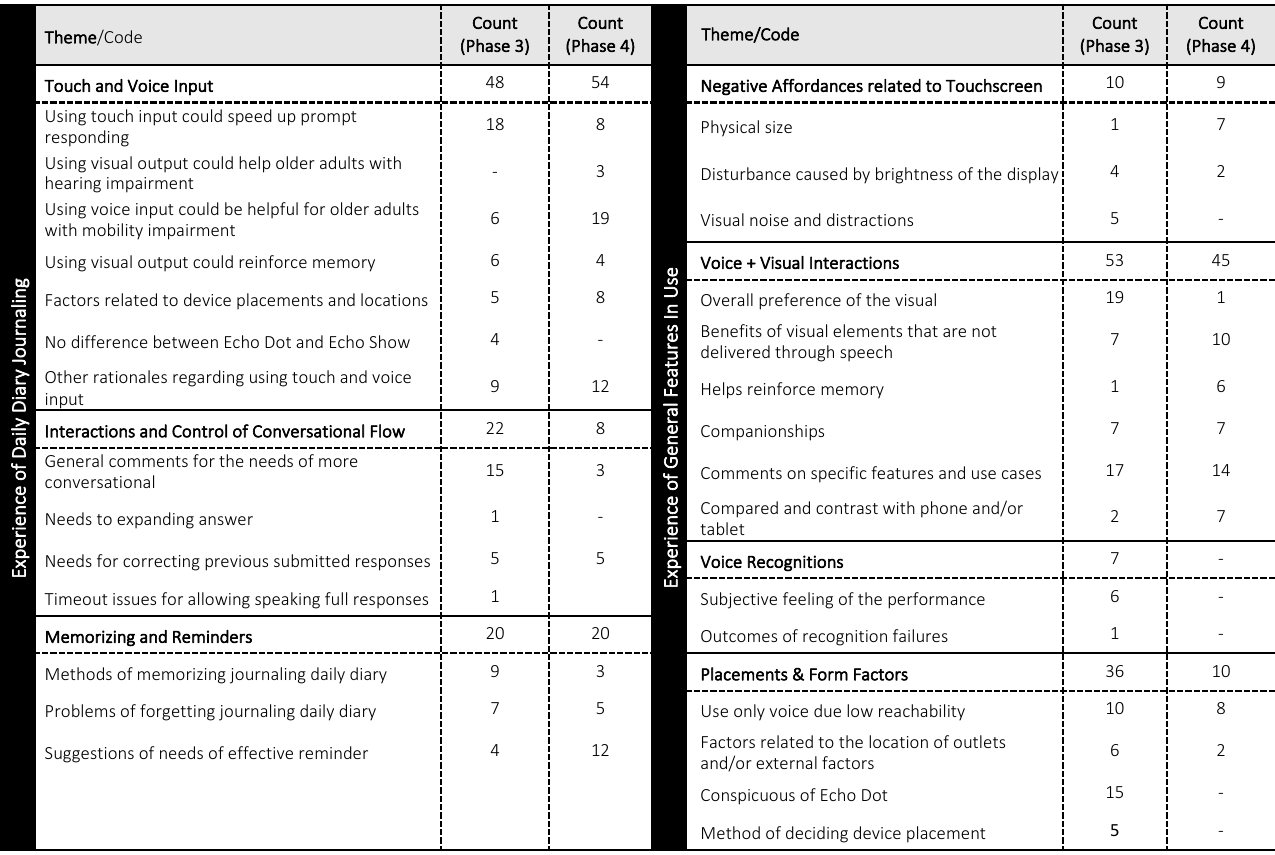}
    \vspace{-0.25in}
    \caption{The codebook resulted from our qualitative analysis, showing three themes (bold) related to the topic of {\it Experience of Daily Diary Journaling}, and another three themes (bold) related to the topic of {\it Experience of General Features In Use}. The ``Count'' refers to the number of participants' quote tagged with corresponding theme (or code). Notably, it is possible that more than one codes are assigned to a specific quote.}
    \Description{Figure 17 shows the codebook resulted from our qualitative analysis, showing three themes (bold) related to the topic of Experience of Daily Diary Journaling, and another three themes (bold) related to the topic of Experience of General Features In Use. The “Count” refers to the number of participants’ quotes tagged with corresponding theme (or code). Notably, it is possible that more than one code is assigned to a specific quote. 29 codes were yielded spanning across three themes (``Touch and Voice Input'', ``Interactions and Control of Conversational Flow'', and ``Memorizing and Reminders'') under the topic of ``Experience of Daily Diary Journaling'' and another three themes (``Negative Affordances related to Touchscreen'', ``Voice + Visual Interactions'', ``Voice Recognition'', and ``Placements & Form Factors'') under the topic of ``Experience of General Features In Use'';}
    \label{fig::app::codebook_diary_gu}
\end{figure*}

Fig.~\ref{fig::app::codebook_device_setups} shows the codebook and themes yielded from qualitative analysis of interview data collected from Phase $2$ of the study.
Similarly, Fig.~\ref{fig::app::codebook_diary_gu} shows the codebook and themes of qualitative analysis results from Phase $3$ and Phase $4$. 

\clearpage
\section{Ethical Disclaimers}\label{sec::app::ethic}
This work has been approved by the Institutional Review Board (IRB).
Before the study, all participants have been introduced and signed the informed consent as well as the video and audio recording consent.
Upon completing the study, the devices were reset and awarded to the participants as incentive, which is around \$$130$ as the time of writing (May 3, 2023) for both Echo Dot and Show. 
During the co-design workshop (Phase~$4$), participants were allowed to disable their camera and/or rename their Zoom account as needed, if they were not comfortable with showing camera feeds and their name to other study participants.
We have gained consents for academic publications of all figures that are involved with anonymous participants.
All analysis data has been unlinked with Personal Identifiable Information~(PII) as per regulated in our IRB, and were stored in a secure cloud storage service, which complies with the Health Insurance Portability and Accountability Act~(HIPAA).
Due to the impacts of COVID-19 during Phase~$2$ of our study that required in-person visit, all research assistants strictly obeyed the guidance and regulations issued by local health authorities (\eg~wearing masks and having a negative COVID-19 PCR test before the visit).

\end{appendices}

\end{document}